\documentclass[11 pt,a4paper]{article}
\usepackage[english]{babel}

\usepackage[top=1in, bottom=1in, left=1in, right=1in]{geometry}
\usepackage[subtle]{savetrees}
\usepackage[compact]{titlesec}

\usepackage[nodisplayskipstretch]{setspace}
\setdisplayskipstretch{0}

\usepackage{placeins}
\usepackage{amssymb}
\usepackage{color}
\usepackage{colortbl}
\usepackage{subcaption}
\usepackage{rotating}
\usepackage{amsmath, amsthm}
\usepackage{nicefrac}
\usepackage{multicol}
\usepackage{multirow}
\usepackage{tabularx}
\usepackage{dsfont}    
\usepackage{bm}
\usepackage{url}
\usepackage{algpseudocode}
\usepackage{algorithm}

\usepackage{footnote}
\usepackage{tablefootnote}

\usepackage{caption}
\makesavenoteenv{algorithm}

\usepackage{bbm}
\usepackage{amsfonts}
\usepackage{graphicx}
\usepackage{psfrag}
\usepackage{fancybox}
\usepackage{booktabs}
\usepackage[flushleft]{threeparttable}
\usepackage{lscape}
\usepackage{ulem}
\usepackage{booktabs}
\usepackage{xcolor}
\usepackage{csquotes}
\usepackage{authblk}
\usepackage[shortlabels]{enumitem}

\usepackage[hidelinks]{hyperref}
\usepackage{xr-hyper}
\graphicspath{ {Figures/} }

\algdef{SE}[SUB]{SubStep}{EndSubStep}[1]{#1}{}
\algtext*{EndSubStep} 

\setenumerate{noitemsep,topsep=0pt,parsep=5pt,partopsep=0pt}
\setitemize{noitemsep,topsep=0pt,parsep=5pt,partopsep=0pt}

\usepackage{natbib}
\setcitestyle{authoryear, aysep={}, yysep={;}}

\newcommand{\bmat}[1]{
  \renewcommand*{\arraystretch}{1.0}
  \begin{bmatrix}#1\end{bmatrix}%
}

\makeatletter
\newcommand{\multiline}[1]{%
  \begin{tabularx}{\dimexpr\linewidth-\ALG@thistlm}[t]{@{}X@{}}
    #1
  \end{tabularx}
}
\makeatother

\DeclareCaptionSubType{algorithm}


\title{Q-approximation of operating characteristics of clinical trial designs}

\author[1]{Susanna Gentile\thanks{Corresponding author: susanna.gentile@uniroma1.it}}
\author[2]{Daniel E. Schwartz}
\author[3]{Riddhiman Saha}
\author[3,4]{Lorenzo Trippa}
\affil[1]{Department of Statistical Sciences, Sapienza University of Rome}
\affil[2]{Massachusetts General Hospital Biostatistics and Harvard Medical School}
\affil[3]{Department of Biostatistics, Harvard T.H. Chan School of Public Health}
\affil[4]{Department of Data Science, Dana-Farber Cancer Institute}

\newcommand{\bne}{\begin{eqnarray}}
\newcommand{\ene}{\end{eqnarray}}
\newcommand{\be}{\begin{eqnarray*}}
\newcommand{\ee}{\end{eqnarray*}}

\newcommand{\E}{\mathbb{E}}

\newtheorem{proposition}{Proposition}
\newtheorem{corollary}{Corollary}
\theoremstyle{definition}

\begin{document}
\date{}
\maketitle
\begin{abstract}

Designing clinical trials requires evaluating multiple operating characteristics (OCs), such as the likelihood of an early stopping decision, the probability of detecting a treatment effect, and the Type I error rate. In most cases, these evaluations are based on computationally intensive Monte Carlo simulations. As the complexity of clinical trials and the use of adaptive designs increase, the computational burden can quickly become prohibitive. We introduce a strategy for rapidly approximating OCs, called the Q-approximation.
Our approach is based on quadratic approximations of the log-likelihood and asymptotic arguments. 
The main idea is to replace simulation of full trial datasets with simulation of the approximate likelihood functions that determine the trial’s interim and final decisions.
The Q-approximation approach can be applied to any trial design that uses data analysis methods coherent with the likelihood principle, including multistage designs with early stopping, adaptively randomized designs, and designs that leverage external data. We illustrate the approach with several examples and show that it provides an accurate approximation of important OCs while reducing the computation time compared to Monte Carlo simulations. In particular, in our experiments, the standard Monte Carlo approximation of OCs requires 150 to 1,900 times greater computing budget than Q-approximations to achieve comparable levels of accuracy.
By enabling fast OC evaluations, Q-approximations can support the broader use of innovative trial designs in both applied trial planning and methodological development. 

\end{abstract}


\section{Introduction}\label{section: Introduction}
 
When investigators design clinical trials, they need to consider multiple operating characteristics (OCs), such as the expected number of enrolled patients in a multistage design, the probability of detecting a positive treatment effect in a randomized clinical trial (RCT), and the overall costs. These OCs quantify critical trade-offs, such as the balance between estimation accuracy and sample size. Importantly, the assessment of OCs is necessary for comparing innovative designs (e.g., adaptive designs) to simpler ones and it supports investigators in the selection of the sample size and other aspects of the trial design, such as the number of interim analyses. Computing OCs is also essential for interactions of the study team with regulatory agencies and other stakeholders. For example, the  Food and Drug Administration recommends rigorous control of false positive results based on interpretable metrics \citep{fda_considerations_2023,fda_adaptive_2019}. 
The lack of fast, reliable tools for evaluating OCs is therefore a major barrier to both the adoption and development of more efficient clinical trial designs \citep{golchi_estimating_2022, han_sensitivity_2024}.

The OCs of a clinical trial depend on several unknown parameters, including the enrollment rate, the distribution of the outcomes under the experimental and control treatments, and pre-treatment covariates in the enrolled population. Researchers typically simulate clinical trials using a comprehensive model with parameters $\omega \in \Omega$ which describe all of these factors. In this manuscript, we refer to a specific parametrization $\omega$ as a \textit{scenario}. Evaluating how OCs vary across plausible scenarios is crucial to choosing an appropriate design. Considering multiple scenarios is also typically beneficial for assessing the robustness of the statistical plan, for example quantifying risks associated with misspecification of the model that will be used for data analyses during and at completion of the trial \citep{white_maximum_1982}. The selection of an adequate set of simulation scenarios has been discussed in the literature; see for example \cite{thabane_tutorial_2013} and  \cite{han_sensitivity_2024}.

Researchers usually rely on Monte Carlo simulations of the clinical trial to compute OCs \citep{robert_monte_2010, chang_monte_2010, morris_using_2019}. However, these simulations can be very time-consuming. For instance, Bayesian adaptive designs often require approximations of posterior probabilities using MCMC algorithms. Therefore, it can be challenging to assess OCs using simulation replicates, and difficulties persist despite the availability of parallel computing and algorithms to efficiently simulate clinical trials \citep{golchi_estimating_2022, golchi_estimating_2024, han_sensitivity_2024, chang_monte_2010, hagar_economical_2025}. 

As an example of this computational burden, consider a Phase 3 RCT with an interim analysis for early futility stopping. The investigators need to choose the sample size, the randomization ratio, and the threshold of a posterior probability that triggers early stopping. With three possible values for each of these features, we have a total of 27 candidate designs. Assume these designs will be compared across only 25 scenarios, defined by 5 values of the treatment effect and 5 accrual rates. Simulating 10,000 trials for each design and scenario, with 1 second of computation per trial, would require over 1,800 computing hours. Although parallel computing can accelerate this process, comprehensive comparisons of trial designs based on Monte Carlo simulations may still be infeasible. As a result, investigators might focus on a narrow set of scenarios (a strategy associated with high risks for the whole drug development process) or evaluate only the simplest statistical designs and methods to reduce the effort necessary for trial simulations.

We propose a likelihood-level simulation framework for approximating
OCs in substantially shorter computation time than 
standard
Monte Carlo simulations. The approximation method that we introduce is based on three considerations:
\begin{description}
    \item[A] In many clinical trial designs the interim decisions and final analyses depend only on the likelihood function. 
    In other words, the statistical plans of these clinical trials adhere to the likelihood principle (see \cite{berger_likelihood_1988} for a formal definition).
    \item[B] Under mild regularity conditions, the log-likelihood can be approximated with a quadratic function.
    \item[C] The distributions of the center and curvature of the quadratic approximation can be derived using asymptotic results \citep[for example,][]{van_der_vaart_asymptotic_2012, white_maximum_1982}.
\end{description}
Together, considerations \textbf{A}, \textbf{B}, and \textbf{C} imply that we can approximate OCs by directly generating approximate likelihood functions, which we call \textit{Q-likelihoods}, instead of simulating  trials as in Monte Carlo studies. We call this strategy the \textit{Q-approximation} of OCs as it arises from a quadratic approximation of the log-likelihood.

We emphasize that the methodological innovation of the Q-approximation is to use classical likelihood approximations as a tool for efficient evaluation of OCs. The approach treats the asymptotic distribution of the likelihood approximation as a simulation model for the quantities that drive the trial’s decisions, including maximum likelihood estimates, p-values, posterior probabilities, and predictive probabilities. This provides a practical, general framework that enables rapid OC evaluations and can be applied across many scenarios and candidate designs.

We provide a simple toy example, which we will refer to as \textbf{Example 1}:

\textit{Design and OC of interest:} Consider a single-arm trial (SAT) with $n = 50$ patients and no interim analyses. The aim is to test whether the response rate $\theta$ exceeds a reference rate of 0.4, using a Bayesian model with binary outcomes $Y_i \overset{iid}{\sim} \mbox{Ber}(\theta)$ and a uniform prior $\theta \sim \mbox{Unif}(0, 1)$. Conditional on the data $D = (Y_1, \dots, Y_n)$ the posterior density of $\theta$, denoted as $\pi_\theta( \cdot \mid D)$, is proportional to the Bernoulli likelihood $L(\theta ; D) =  \theta^{Y_{\boldsymbol{\cdot}}} (1 - \theta)^{n -  Y_{\boldsymbol{\cdot}}}$ where $Y_{\boldsymbol{\cdot}} = \sum_{i=1}^n Y_i$. We use the notation $\pi_\theta(\cdot \mid D)$ because later we will discuss the posteriors of quantities other than $\theta$.  The trial reports a positive result if at the end of the study $\Pi_\theta([0.4, 1] \mid D) > 0.9$, where $\Pi_\theta(B \mid D)$ denotes the posterior probability of an event $B$. The OC of interest in this example is the probability of reporting a positive result.

\textit{Monte Carlo approximation:} Assuming a response rate of 0.5 (our scenario $\omega$), to approximate the probability of a positive finding, we need to iterate the following two steps $R$ times: (i) generate the data $D = (Y_1, \ldots, Y_n),$ where $\;Y_i \sim \mbox{Ber}(0.5)$, for $i = 1, \dots, n$ and (ii) evaluate if $\Pi_\theta([0.4,1] \mid D) > 0.9$.  

\textit{Q-Approximation:} Alternatively, we can approximate the probability of a positive finding using considerations \textbf{A, B} and \textbf{C}. In particular:
\begin{itemize}

\item[\textbf{A}] The trial follows the likelihood principle since the primary analysis is based on $\pi_\theta(\theta \mid D) \propto L(\theta; D)$.

\item[\textbf{B}] We can approximate the likelihood function $L(\cdot;D)$ with a Gaussian density (i.e., a quadratic function after logarithmic transformation)
$$
\widetilde{L} \left(\theta; \mu, \Sigma^{-1} \right) \propto \exp \left\{ - \frac{1}{2} \Sigma^{-1}(\theta - \mu)^2 \right\},
$$
where $\mu = \hat{\theta} = \frac{Y_{\boldsymbol{\cdot}}}{n} $ is the maximum likelihood estimator (MLE) of $\theta$, and the precision $\Sigma^{-1} = J \left( \hat{\theta} \right) =  \frac{n}{\hat{\theta} (1 - \hat{\theta})}$ is the observed Fisher information. Thus, since we use a uniform prior on $\theta$, the posterior density of $\theta$ is approximately $\widetilde{\pi}_\theta \left( \cdot ; \mu, \Sigma^{-1} \right)  = \phi(\cdot; \mu, \Sigma)$,
where $\phi(\cdot; a, b)$ is a univariate Gaussian density with mean $a$ and variance $b$.

\item[\textbf{C}] The sampling distribution of $\hat{\theta}$ and $J \left( \hat{\theta} \right)$ can be approximated using standard asymptotic results. Thus, we can approximately mimic a trial likelihood $L(\theta; D)$ with a Q-likelihood $\widetilde{L}(\theta; C, V)$, where the center $C$ and curvature $V$ are sampled based on standard asymptotic results for $\hat{\theta}$ and $J \left( \hat{\theta} \right)$. 

\end{itemize}

To approximate the OC of interest, we can repeat the following two steps $R$ times: (i) simulate $(C,V)$ based on standard asymptotic results on the MLE, and (ii) evaluate at each iteration whether or not $\widetilde{\Pi}_\theta \left( [0.4, 1] ; C, V \right) > 0.9$. Here $\widetilde{\Pi}_\theta \left(B; C, V \right)$ denotes the probability of an event $B$ (i.e. subset of the $\theta$ parameter space) under the ``Q-posterior'' distribution obtained by multyplying the Guassian Q-likelihood $\widetilde{L}\left(\theta; C, V \right)$ and the uniform prior on $\theta$. This Q-approximation strategy avoids generating the dataset $D$ by directly simulating the location $C$ and curvature $V$ of the likelihood approximation. 

In this paper, we will show how to use  Q-approximations  to investigate the OCs of more complex trial designs, and discuss examples where the use of Monte Carlo simulations become computationally prohibitive. In particular, we will focus on adaptive designs and trial designs that leverage external data. Adaptive designs include interim analyses for major decisions based on the available data, such as early stopping \citep{pocock_group_1977, gsponer_practical_2014}, increasing the sample size \citep{gould_sample_2001, schmidli_robust_2014}, or updating the randomization probabilities \citep{simon_adaptive_1977, thall_practical_2007, robertson_response-adaptive_2023}. Also, several trial designs that integrate external control data have been proposed. For example, \cite{kotecha_leveraging_2024} compared methods that combine RCT data and external data to estimate treatment effects, while \cite{ventz_use_2022} proposed leveraging external datasets for interim decisions. 

In addition to \textbf{Example 1}, in the following sections we will discuss several  examples to illustrate the broad applicability of Q-approximations. \textbf{Example 2} is a RCT with binary outcomes. \textbf{Example 3} is a trial design that uses external data for interim decisions  \citep{ventz_use_2022}. Finally, \textbf{Example 4} is a trial design with Bayesian adaptive randomization \citep{lee_bayesian_2010}.

\section{Method}

We propose a procedure to approximate the OCs of  clinical trials. Similarly to a Monte Carlo simulation study, the inputs are the trial protocol (i.e., a detailed description of the design and data analysis plan) and the simulation scenarios. The outputs are the OCs' approximations. In this section, after introducing the notation (Table \ref{Tab:notation}), we describe our Q-approximation procedure starting from single stage designs. We will initially use our Example 1 to illustrate the methodology.  We will then extend the presentation of Q-approximations  to more complex designs.

\subsection{Notation}

\textbf{Protocol of the clinical trial:} 
The protocol $\mathcal{T}$ provides details on the design of the trial and the data analysis plan, including the outcomes and covariates that will be measured, the sample size, details on the randomization procedure and interim decisions, and the procedures for data analysis.
The dataset at the end of the study is $D = (Y_i, X_i, A_i)_{i \leq n}$, where $Y_i$ is the primary outcome for patient $i$, $X_i$ is a vector of covariates, $A_i$ is the treatment arm, and $n$ is the  sample size.

All interim and final data analyses for the trial are based on a model $p_{\theta}(Y_i \mid X_i, A_i)$, indexed by a parameter $\theta \in \Theta \subseteq \mathbb{R}^d$. The data analysis plan assumes conditionally independent observations, and the likelihood function is $L(\theta ; D) = \prod_{i = 1}^n p_\theta(Y_i \mid X_i, A_i)$. We use $\ell(\theta ; D)$ to indicate the log-likelihood. The maximum likelihood estimator (MLE) of $\theta$ is $\hat{\theta}$. If the analysis is Bayesian, then the protocol will specify the prior density for $\theta$, denoted by $\pi_\theta (\cdot)$. 
Also, $\pi_\theta(\cdot \mid D)$ denotes the posterior density of $\theta$ given $D$, and $\Pi_\theta(B \mid D)$ is the posterior probability of a  set $B \subseteq \Theta$. As needed, we change the subscripts to denote posteriors for parameters other than $\theta$; for example, $\pi_\delta (\cdot \mid D)$ will denote the posterior density of $\delta$, a transformation  of $\theta$.

The aim of the trial is to infer a parameter $\delta$. 
For example $\delta$ might be the difference between the mean outcomes of the two arms in an RCT. We consider only data analysis plans and interim decisions that adhere to the likelihood principle \citep{berger_likelihood_1988}. In other words, the likelihood function is sufficient for decision-making. 

\textbf{Scenarios:} We evaluate the OCs of a trial design.  The distribution of the  data is $D \sim g_{\omega, \mathcal{T}}$. Here, the distribution $g_{\omega,\mathcal{T}}$ is indexed by the trial protocol $\mathcal{T}$ and a parameter $\omega \in \Omega$. The scenario $\omega$ includes all aspects of the data distribution that are not controlled by the investigator, such as the treatment effects and the enrollment rate. Also, we denote  by $q_{\omega}(Y_i \mid X^+_i, A_i)$ the sampling distribution of the individual outcomes conditional on covariates $X^+_i$ and treatment $A_i$. In some cases, the covariates $X^+_i$ used in the simulation model $g_{\omega, \mathcal{T}}$ can differ from the covariates $X_i$ used in the analysis model $p_\theta$. For example $X_i$ might be a sub-vector of $X^+_i$. We also emphasize that $q_\omega$ in some cases will not belong to the class of distributions $\left(p_\theta; \theta \in \Theta\right)$. That is, the protocol's analysis model could be misspecified.

\begin{table}[!htbp]
\small
\renewcommand{\arraystretch}{1.25} 
\centering
\caption{Notation}\label{Tab:notation}

\begin{tabular}{p{0.6\textwidth}p{0.4\textwidth}}

\toprule
Protocol & $\mathcal{T}$  \\
Data at completion of the study & $D = \left(Y_i,\, X_i, A_i\right)_{i \leq n}$ \\
Sample size & $n$ \\
Parameters of the data analysis model & $\theta \in \Theta$ \\
Parameters of the simulation scenario & $\omega\in \Omega$ \\

Conditional distribution of $Y_i$ (analysis model) & $ p_\theta\left(Y_i \mid  X_i, A_i\right)$\\
Conditional distribution of $Y_i$ (simulation model) & $ q_\omega\left(Y_i \mid X^+_i,A_i\right)$ \\ 
Distribution of the  dataset $D$ (simulation model) & $g_{\omega, \mathcal{T}}$ \\
Likelihood and log-likelihood  & $L\left(\theta;D\right)$ and $ \ell\left(\theta;D\right)$ \\
Operating characteristic of interest & $\psi = E_{g_{\omega, \mathcal{T}}} \left[ h(D) \right] = E_{g_{\omega, \mathcal{T}}} \left[ h' \left( L \left( \cdot; D \right) \right) \right]$ \\
Maximum likelihood estimator & $\hat{\theta}$\\
Observed Fisher information  & $J = J \left( \hat{\theta} \right)$ \\
 Likelihood approximation with center $\hat{\theta}$ and curvature $J$ 
& $\widetilde{L} \left( \theta; \hat{\theta}, J \right)$ \\
$d$-dimension Gaussian distribution with mean $\mu$ and variance $\Sigma$ & $N_d \left( \mu, \Sigma \right)$ \\
$d$-dimensional Gaussian density with mean $\mu$ and variance $\Sigma$ & $\phi_d\left(\cdot; \mu, \Sigma\right)$ \\
$d$-dimensional Gaussian CDF with mean $\mu$ and variance $\Sigma$ & $\Phi_d\left(\cdot; \mu, \Sigma\right)$ \\
KL projection of $q_\omega$ onto $\left(p_\theta; \theta \in \Theta\right)$  & $\theta^*$ \\
Asymptotic variance of the MLE 
& $\mathcal{V}^*$  \\
Expected negative Hessian of $\ell \left( \theta; \left(Y_i, X_i, A_i\right) \right)$ (one observation) at $\theta^*$ & $\mathcal{J}^*$ \\
Q-likelihood with random center $C$ and fixed curvature $V$ & $\widetilde{L}\left(\theta; C, V\right)$ \\
Random vector sampled from the asymptotic distribution of $\hat{\theta}$
& $C$ \\
Asymptotic approximation of the curvature of $\ell\left(\theta; D\right)$  & $V$
\\
Approximate sampling distribution of $\hat{\theta}$ based on asymptotics & $f_{\omega, \mathcal{T}}$ \\
Prior and posterior density of $\theta$ & $\pi_\theta\left(\cdot\right), \; \pi_\theta\left(\cdot \mid D\right)$ \\
Posterior probability (measure) of a set $B \subseteq \Theta$ & $\Pi_\theta\left(B \mid D\right)$ \\
Q-posterior density implied by $\widetilde{L}\left(\theta; C, V\right)$
& $\widetilde{\pi}_\theta\left(\cdot; C^{(P)}, V^{(P)}\right)$  \\
Q-posterior probability (measure) of a set $B \subseteq \Theta$
& $\widetilde{\Pi}_\theta\left(B; C^{(P)}, V^{(P)}\right)$  \\
Center of the Q-posterior &  $C^{(P)}$ \\
Curvature of the Q-posterior & $V^{(P)}$ \\
\midrule
Monte Carlo or Q-approximation replicates & $r = 1, \dots, R$ \\
\midrule
Total number of stages of a clinical trial & $S$\\
Stage-specific data and cumulative data at stage  $s$ & $D_{s}$ and $ D_{1:s}$\\
Stage-specific and cumulative likelihoods of $D_{s}$ and $D_{1:s}$ & $L(\theta; D_{s})$ and $L(\theta;D_{1:s})$ \\
Stage-specific and cumulative Q-likelihoods with random centers and curvatures
& $\widetilde{L}\left(\theta; C_{s}, V_{s}\right)$ and $\widetilde{L}\left(\theta; C_{ 1:s},V_{ 1:s}\right)$ \\

\bottomrule
\end{tabular}

\end{table}

\textbf{The OCs of interest:}
We consider OCs that can be expressed as expectations $\psi = \E_{g_{\omega, \mathcal{T}}}[h(D)]$, where $h$ is a function that maps datasets to the real line and $\E_{g_{\omega, \mathcal{T}}}$ indicates expectation with respect to $g_{\omega, \mathcal{T}}$. In Example 1 (see Section \ref{section: Introduction}) the OC of interest is the probability of reporting evidence of a response rate greater than 0.4. The function $h$ in this case is: 
\bne\label{eq:condition_binary_SAT}
h(D) = \mathbbm{1}\left\{\Pi_\theta([0.4, 1] \mid D )>0.9\right\} = \mathbbm{1}\left\{\left [1 - \mbox{Beta}\left(0.4; \phantom{i} Y_{\boldsymbol{.}} + 1, n - Y_{\boldsymbol{.}} + 1\right)\right ] >0.9\right\},
\ene
where 
$\mathbbm{1}\{\cdot\}$ denotes the indicator function, $\text{Beta}(\cdot; a, b)$ is the cumulative distribution function (CDF) of a Beta distribution with parameters $a$ and $b$, and $Y_{\boldsymbol{.}} = \sum_{i = 1}^n Y_i$.

\subsection{Q-approximation of OCs} \label{section:single_stage_Q}

We consider a clinical trial without interim analyses and describe how the likelihood principle (\textbf{Consideration A}), a quadratic approximation of the log-likelihood (\textbf{Consideration B}), and  asymptotic results (\textbf{Consideration C}) can be combined to approximate OCs. 

The estimation of OCs usually relies on Monte Carlo (MC) simulations of the trial \citep{chang_monte_2010, morris_using_2019}.  The MC-approximation of an OC $\psi = \E_{g_{\omega, \mathcal{T}}}[h(D)]$ is 
\bne \label{eq:mc_standard}
\hat{\psi}_{MC} = \frac{1}{R}\sum_{r = 1}^R h\left(D^{(r)}\right),
\ene
where $D^{(r)} \overset{iid}{\sim} g_{\omega, \mathcal{T}}$,  $r = 1, \dots, R$, are  simulated datasets.   Under mild assumptions, the law of large numbers implies that $\hat{\psi}_{MC} \overset{p}{\to} \psi$ as the number of replicates $R \to \infty$. Computing $h\left(D^{(r)}\right)$ can be slow, for example if the data analysis involves complex Bayesian models. Furthermore, generating datasets $D^{(r)}$ from the distribution $g_{\omega, \mathcal{T}}$ can be time-consuming. We propose to accelerate the approximation of OCs by leveraging \textbf{Considerations A}, \textbf{B} and \textbf{C}.

\textbf{Consideration A:} The likelihood principle implies that there exists a function $h'$ such that $h(D) = h'(L(\cdot; D))$ for every dataset $D$ in the support of $g_{\omega, \mathcal{T}}$. Here $h'$ is a map with domain equal to the space of  functions from $\Theta$ to the non-negative real numbers. As a result, the OC can be rewritten as $\psi = E_{g_{\omega, \mathcal{T}}} \left[ h' \left( L(\cdot; D) \right) \right]$ and the MC-approximation \eqref{eq:mc_standard}  as
\bne\label{eq:mc_lik}
\hat{\psi}_{MC} = \frac{1}{R}\sum_{r = 1}^R h'\left(L\left(\cdot; D^{(r)}\right)\right).
\ene 

In Example 1,  $h'$ transform the likelihood function $L\left(\cdot;D^{(r)}\right)$ into a point, either $\{0\}$ or $\{1\}$.  The function $h'$ indicates whether or not the posterior probability of a positive treatment effect exceeds 0.9. 

\textbf{Consideration B:}  Under mild regularity conditions, we can approximate the log-likelihood $\ell\left(\theta;D\right)$ with a quadratic function. Specifically, a second-order Taylor expansion around the MLE $\hat{\theta}$ gives
\be
\ell\left(\theta;D\right) \approx  \ell\left(\hat{\theta}; D\right) - \frac{1}{2}\left(\theta - \hat{\theta}\right)^{\top}  J\left(\hat{\theta}\right) \left(\theta - \hat{\theta}\right), 
\ee
where $J \left(\hat{\theta} \right) = - \frac{\partial^2}{\partial \theta \partial \theta^\top} \ell(\theta;D)\big|_{\theta = \hat{\theta}}$ is the observed Fisher information of the analysis model $\ell(\theta; D)$. 
Throughout the paper we use the notation $J = J \left(\hat{\theta} \right)$, except when the Fisher information is evaluated at another point. This quadratic approximation implies that, under mild regularity conditions and up to a multiplicative constant, the likelihood $L(\theta; D)$ can be approximated by the Gaussian density
$$ \widetilde{L} \left(\cdot; \hat{\theta}, J \right) = \phi_d \left(\cdot; \hat{\theta}, J^{-1} \right). $$
Here $\phi_d \left( \cdot; \mu, \Sigma \right)$ denotes  a $d$-dimensional Gaussian density  with mean  $\mu$ and variance $\Sigma$. 

Since $\widetilde{L}$ approximates $L$ we can approximate  $h(D) = h' \left( L(\cdot; D) \right)$  with $h' \left( \widetilde{L}\left( \cdot; \hat{\theta}, J \right) \right)$.  In particular, replacing $L \left( \cdot ; D^{(r)} \right)$ with $\widetilde{L} \left( \cdot; \hat{\theta}^{(r)}, J^{(r)} \right)$  in the right-hand side of equation \eqref{eq:mc_lik}, we have
\bne\label{eq:mc_q_approx_sample}
\hat{\psi}_{MC} \approx
\frac{1}{R}\sum_{r = 1}^R h'\left(\widetilde{L}\left(\cdot;\hat{\theta}^{(r)}, J^{(r)} \right)\right), 
\ene
where $\hat{\theta}^{(r)}$ and $J^{(r)}  = - \frac{\partial^2}{\partial \theta \partial \theta^\top} \ell \left(\theta; D^{(r)} \right)\big|_{\theta = \hat{\theta}^{(r)}}$ are the MLE and the observed Fisher information.

In Example 1 $\hat{\theta}^{(r)} = \overline{y}^{(r)}$, where $\overline{y}^{(r)} = \frac{1}{n}\sum_{i = 1}^n Y_i^{(r)}$ denotes the sample mean, and $J^{(r)} = \frac{n}{\hat{\theta}^{(r)}\left(1 -\hat{\theta}^{(r)}\right)}$. The approximation of the likelihood is $\widetilde{L}\left( \cdot; \hat{\theta}^{(r)},  J^{(r)} \right) = \phi\left(\cdot; \overline{y}^{(r)}, \frac{\overline{y}^{(r)}(1 - \overline{y}^{(r)})}{n} \right)$. Applying the map $h'$ to the function  $\widetilde{L}\left(\cdot; \hat{\theta}^{(r)}, J^{(r)} \right)$ we obtain an approximate version of equation  \eqref{eq:condition_binary_SAT}: 
$$
h'\left(\widetilde{L}\left( \cdot; \hat{\theta}^{(r)}, J^{(r)} \right)\right) = \mathbbm{1}\left\{\left [1 - \Phi\left(0.4; \overline{y}^{(r)}, \frac{\overline{y}^{(r)}(1 - \overline{y}^{(r)})}{n} \right) \right ]\geq 0.9\right\},
$$
where $\Phi(\cdot; a, b)$ is the CDF of a univariate Gaussian distribution with mean $a$ and variance $b$.

\textbf{Consideration C:}
The approximation \eqref{eq:mc_q_approx_sample} requires simulating independent datasets $D^{(r)}$ and computing their MLEs $\hat{\theta}^{(r)}$ and curvatures  $J^{(r)}$. Our Q-approximation framework does not involve simulating $D^{(r)}$, but instead directly generates independent functions  $\widetilde{L} \left(\cdot; C^{(r)}, V \right)$, for $r = 1, \dots, R$. The centers $C^{(r)} \overset{iid}{\sim} f_{\omega, \mathcal{T}}$ are sampled from an approximation $f_{\omega, \mathcal{T}}$ of the sampling distribution of $\hat{\theta}$ based on standard asymptotic arguments. The curvature $V$ is fixed, again  on the basis of standard asymptotic results. The Q-approximation of the OC $\psi$ is 
\bne \label{eq:q_approximation}
\hat{\psi}_{Q} = \frac{1}{R}\sum_{r = 1}^R h' \left( \widetilde{L}  \left(\cdot; C^{(r)}, V \right) \right).
\ene
This approximation is closely related to expression \eqref{eq:mc_q_approx_sample}. Under mild regularity conditions the following properties hold as $n \to \infty$:
\begin{enumerate}[(a)]
    \item $\hat{\theta} \overset{p}{\to} \theta^*$ for some parameter value $\theta^* \in \Theta$,
    
    \item $\sqrt{n} \left( \hat{\theta} - \theta^* \right) \overset{d}{\to} N_d \left( 0, \mathcal{V}^* \right)$ for some positive semi-definite matrix $\mathcal{V}^*$, and
    
    \item $n^{-1} J \overset{p}{\to} \mathcal{J}^*$ for some positive semi-definite matrix $\mathcal{J}^*$.
\end{enumerate}
Properties (a) and (b) justify replacing $\hat{\theta}^{(r)}$ in expression \eqref{eq:mc_q_approx_sample} with $C^{(r)} \overset{iid}{\sim} f_{\omega, \mathcal{T}} = N \left( \theta^*, n^{-1}\mathcal{V}^* \right)$, and property (c) justifies replacing $J^{(r)}$ with $V = n \mathcal{J}^*$.

We emphasize that the function $\widetilde{L} \left(\cdot; C^{(r)}, V \right)$ is \textit{not} linked to a dataset $D^{(r)}$, and we refer to it as the  \textit{Q-likelihood}. Indeed, the center $C^{(r)}$ is \textit{not} the MLE for a specific dataset $D^{(r)}$; it is just a random variable with distribution similar to $\hat{\theta}^{(r)}$. The curvature $V$ is not the observed Fisher information for a specific dataset; it is an asymptotic approximation of $J$. Importantly, standard asymptotic arguments imply that $\widetilde{L} \left(\cdot; C^{(r)}, V \right)$ in expression \eqref{eq:q_approximation} and $\widetilde{L} \left( \cdot; \hspace{0.1em} \hat{\theta}^{(r)}, J^{(r)} \right)$ in expression \eqref{eq:mc_q_approx_sample} have similar distributions.

In Example 1, the simulation model is $Y_i \overset{iid}{\sim} Ber(\omega)$ and the asymptotic distribution of $\hat{\theta}$ is $\sqrt{n} \left( \hat{\theta} - \omega \right) \overset{d}{\to} N \Big( 0, \phantom, \omega (1 - \omega) \Big)$ as $n \to \infty$; therefore we sample the centers $C^{(r)}$ from the distribution $f_{\omega, \mathcal{T}} = N \left( \omega, \phantom,  \frac{\omega (1 - \omega)}{n} \right)$. In addition, the observed Fisher information satisfies $n^{-1} J \overset{p}{\to} \frac{1}{\omega (1 - \omega)}$; therefore we set the curvatures of the Q-likelihood as $V = \frac{n}{\omega (1-\omega)}$. Then we compute 
\begin{equation*} \label{eq:ex1_h_tilde}
h'\left(\widetilde{L}\left(\cdot; C^{(r)}, V \right)\right) = \mathbbm{1}\left\{1 - \Phi\left(0.4; C^{(r)}, V^{-1} \right) \geq 0.9\right\}.
\end{equation*}
Algorithms \ref{alg:MC_ex1} and \ref{alg:Q_ex1} summarize the key steps and differences in the MC- and Q-approximations of the power.

\begin{algorithm}[t]

\caption{MC-approximation and Q-approximation, Example 1. OC: probability of reporting a positive result.}
    
\vspace{0.5em}

\begin{subalgorithm}[H]{\textwidth}
    
    \caption{\textbf{MC-approximation}} \label{alg:MC_ex1}
    
    \textbf{Input:} $n$, $\omega$
    \begin{algorithmic}[1]
    \For {replicate $r$ in $1$:$R$}

    \State Sample $D^{(r)} = \{Y_i\}_{i\leq n}$, where $Y_i \overset{iid}\sim \mbox{Ber}(\omega)$.

    \State Compute $h\left(L\left(\cdot; D^{(r)}\right)\right) =  \mathbbm{1}\left\{[1 - \mbox{Beta}\left(0.4; \phantom{i} Y_{\boldsymbol{.}} + 1, n - Y_{\boldsymbol{.}} + 1\right)] \geq 0.9\right\}$.

    \EndFor
    \end{algorithmic}

    \textbf{Output:} $\hat{\psi}_{MC} = \frac{1}{R} \sum_{r = 1}^R h\left(L\left(\cdot; D^{(r)}\right)\right)
    $
\end{subalgorithm}

\vspace{1em}

\begin{subalgorithm}[H]{\textwidth}
    
    \caption{\textbf{Q-approximation}} \label{alg:Q_ex1}
    
    \textbf{Input:} $n$, $\omega$
    \begin{algorithmic}[1]
    
    \For {replicate $r$ in $1$:$R$}

    \State Generate the Q-likelihood $\widetilde{L} \left( \cdot; C^{(r)}, V \right)$, where $C^{(r)} \sim N\left(\omega, \frac{\omega(1 - \omega)}{n} \right)$ and $V = \frac{n}{\omega(1 - \omega)}$.

    \State Compute $h'\left(\widetilde{L}\left( \cdot ; C^{(r)}, V \right) \right) = \mathbbm{1} \left\{  \left[1 - \Phi \left( 0.4; C^{(r)},  V^{-1} \right) \right] \geq 0.9 \right\}$.

    \EndFor
    \end{algorithmic}

    \textbf{Output:} $\hat{\psi}_Q = \frac{1}{R} \sum_{r = 1}^R h' \left(\widetilde{L}\left(\cdot; C^{(r)}, V \right) \right)$
\end{subalgorithm}

\end{algorithm}

In more general settings,
beyond Example 1, $f_{\omega, \mathcal{T}}$ and $V$ can be derived using standard asymptotic results even when the analysis model is misspecified \citep{van_der_vaart_asymptotic_2012, white_maximum_1982}. To derive $f_{\omega, \mathcal{T}}$, first note that the MLE $\hat{\theta}$ is asymptotically normal under mild regularity conditions (see Theorem 3.2 in \cite{white_maximum_1982}). In particular, as the sample size $n \to \infty$,
\bne \label{eq:mle_distribution}
\sqrt{n}(\hat{\theta} - \theta^*) \overset{d}{\rightarrow} N_d \left( 0, \mathcal{V}^* \right);
\ene 
therefore $C \sim f_{\omega, \mathcal{T}} = N_d \left( \theta^*, n^{-1} \mathcal{V}^* \right)$. 
Here $\theta^* = \text{argmin}_{\theta \in \Theta} \E_{g_{\omega, \mathcal{T}}}\left[ D_{KL} \left(q_\omega(\cdot \mid  X_i^+, A_i) \phantom| || \phantom| p_\theta(\cdot \mid  X_i, A_i) \right) \right]$ is the parameter value that minimizes the expected Kullback-Leibler (KL) divergence between the analysis model $p_\theta(\cdot \mid X_i, A_i)$ and the simulation model $q_\omega(\cdot \mid X_i^+, A_i)$.
We remind the reader that $X_i$ are the covariates used in the analysis model, which are a function (e.g., a subvector) of the covariates $X_i^+$ in the simulation model. 
Also, the divergence $D_{KL} \left( q_\omega(\cdot \mid  X_i^+, A_i) \phantom| || \phantom| p_\theta(\cdot \mid  X_i, A_i) \right)$ is a function of the random variables $(X_i^+, X_i, A_i)$ and the expectation $\E_{g_{\omega, \mathcal{T}}} \left[ \cdot \right]$ is taken with respect to their joint distribution under $g_{\omega, \mathcal{T}}$. We emphasize that in this subsection we are considering a simulation model $g_{\omega, \mathcal{T}}$ in which the triplets $(X_i^+, Y_i, A_i)$ are independent and identically distributed; later in the manuscript we will consider adaptive designs where this assumption does not hold.

The asymptotic variance of $\hat{\theta}$ in \eqref{eq:mle_distribution} is $\mathcal{V}^* = \mathcal{J}(\theta^*)^{-1} \mathcal{I}(\theta^*) \mathcal{J}(\theta^*)^{-1}$, where  the function
$\mathcal{J}(\theta) = -\E_{g_{\omega,\mathcal{T}}}\left[\frac{\partial^2}{\partial \theta\partial \theta^\top}\log p_\theta(Y_i\mid X_i,A_i)\right]$ is the expected curvature of the analysis model at an arbitrary point $\theta$, and similarly the function $\mathcal{I}(\theta) = -\E_{g_{\omega,\mathcal{T}}}\left[ \frac{\partial}{\partial \theta} \log p_\theta\left(Y_i\mid X_i, A_i \right)  \left( \frac{\partial}{\partial \theta} \log p_{\theta}\left(Y_i\mid X_i, A_i \right) \right)^\top \right]$ is the expected Fisher information of the analysis model at an arbitrary point $\theta$ \citep{white_maximum_1982}. If the analysis model is correctly specified, i.e., $X_i = X_i^+$ and there exists $\theta \in \Theta$ such that $p_{\theta}(\cdot \mid X_i, A_i) = q_\omega(\cdot \mid X_i^+, A_i)$ for all $X_i, A_i$, then the analytic expression
of $\mathcal{V}^*$
simplifies because $\theta^* = \theta$,   $\mathcal{J}(\theta) = \mathcal{I}(\theta)$, and therefore  $\mathcal{V}^* = \mathcal{J}(\theta^*)^{-1}$. We also note that under mild conditions the observed Fisher information converges  to the expected information, i.e., $n^{-1} J(\hat\theta) \overset{p}{\to} \mathcal{J}^*$ as $n \to \infty$ (by the law of large numbers), so we can set $V = n \mathcal{J}^*$ in equation \eqref{eq:q_approximation}.

\subsection{Graphing the key steps of the Q-approximation procedure} \label{section:mickey_mouse}

\begin{figure}[t]
\centering

\includegraphics[scale = 0.55]{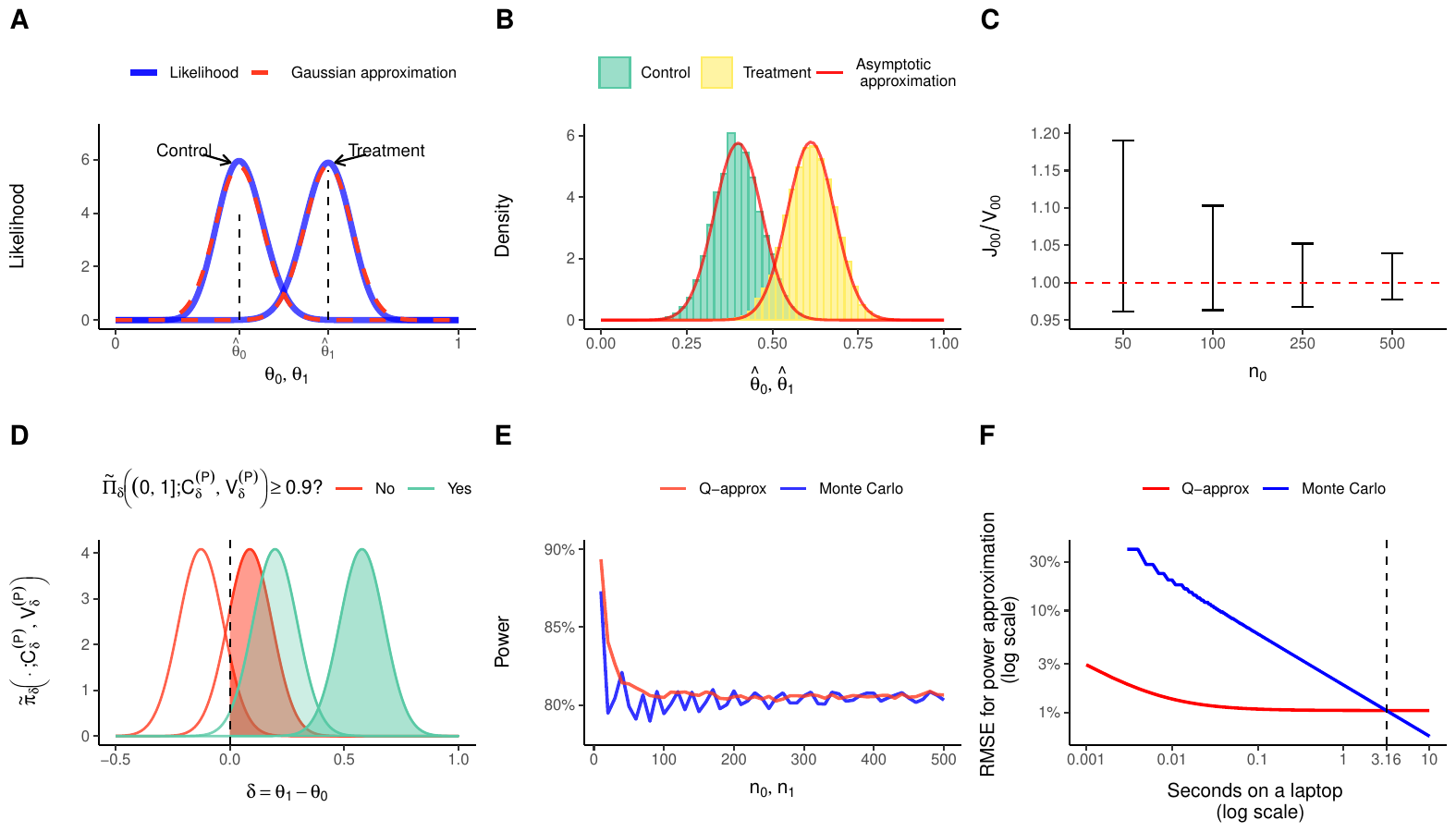}
\caption{Summary of the Q-approximation for Example 2, $n = 100$, $n_0 = n_1 = 50$, and $\omega_{100} = (0.4, 0.61)$. 
\textbf{Panel A} ~shows, for a single dataset, the similarity  between the likelihood $L(\theta; D)$ and its Gaussian approximation $\widetilde{L} \left( \theta; \hat{\theta}, J  \right)$ from equation \eqref{eq:gaussian_lik_ex2}. For the control group, the blue curve is  $L \left( (\cdot, \hat{\theta}_1); D \right)$,  a function of $\theta_0$, and the red curve is the approximation $\widetilde{L} \left( ( \cdot,\hat{\theta}_1); \hat{\theta}, J \right)$.
\textbf{Panel B} ~compares the distribution of the MLE $(\hat{\theta}_0, \hat{\theta}_1)$ across 1,000 datasets $D^{(r)} \overset{iid}{\sim} g_{\omega_{100}, \mathcal{T}}$  and the asymptotic approximation $f_{\omega_{100}, \mathcal{T}}$ used to sample $C^{(r)}$ (alg  \ref{alg:Q_ex2}, line 2). 
\textbf{Panel C} ~shows the sampling distribution of the ratio between the observed information $J_{0,0}$ (from equation~\eqref{eq:gaussian_lik_ex2}) and the fixed curvature $V_{0,0} = n_0 \mathcal{J}_{0,0} = \frac{n_0}{\omega_0 ( 1 - \omega_0 )}$ 
(the first entry of the $V$ matrix in Algorithm~\ref{alg:Q_ex2}, line 2), as $n_0$ increases. Bars show the 5\% and 95\% quantiles.  
\textbf{Panel D} ~ includes four replicates of the function $h'\left( \widetilde{L} \left( \cdot; C^{(r,P)}, V^{(P)} \right) \right) = \mathbbm{1}\left\{\widetilde{\Pi}_\delta \left( (0, \infty]; C_\delta^{(r,P)}, V_\delta^{(P)} \right) > 0.9 \right\}$ (alg  \ref{alg:Q_ex2}, line 3). 
\textbf{Panel E} ~illustrates results of the Q-approximation (red, Algorithm \ref{alg:Q_ex2}) and the MC-approximation (blue, Algorithm \ref{alg:MC_ex2}) for the sequence of scenarios $\omega_n = (0.4, 0.4 + \frac{2.1}{\sqrt{n}})$, with $R = 100,000$ and $M = 20,000$ draws from the posterior in Algorithm \ref{alg:MC_ex2}.
\textbf{Panel F} ~compares the RMSE of $\hat{\psi}_{MC}$ (Algorithm \ref{alg:MC_ex2}, with M = 20,000) with $\hat{\psi}_{Q}$ (Algorithm \ref{alg:Q_ex2}). It shows  the decrease of the RMSE as a function of the time budget (in seconds). Approximations used as many replicates as possible for a given time budget. Since one MC replicate requires approximately 2.1 milliseconds, the RMSE for the MC estimator cannot be computed for the smallest time budgets.}  
\label{fig:ex2}
\end{figure}

In this subsection we illustrate graphically (Figure \ref{fig:ex2}) the key steps of Q-approximations. We focus on a simple RCT with binary outcomes, our Example 2. For ease of presentation, we keep the protocol $\mathcal{T}$ as simple as possible. More complex examples are discussed in later sections.

\textbf{Trial protocol:} 
The data are $D = (Y_i, A_i)_{i \leq n}$, where $n = n_1 + n_0$ and $n_1, n_0$ are the number of patients assigned to the treatment and control arms, respectively. The  design is balanced with $n_1 = n_0$, and the data are analyzed using a  Bayesian model:
\begin{equation} \label{model:ex2_analysis}
    Y_i \mid \theta, A_i = k  \overset{iid}{\sim} \text{Ber}\left(\theta_{k}\right), \quad \theta_k \overset{ind.}{\sim} \text{Beta}\left(\alpha_k, \beta_k\right), \quad \text{for } i = 1,\ldots,n, \text{ and } k = 0, 1,
\end{equation}
where $\theta = (\theta_0, \theta_1)$ denotes the response probabilities under the control ($k = 0$) and experimental ($k = 1$) arms. 
We let $\Pi_\theta(\cdot \mid D)$ denote the posterior distribution of $\theta$, while the posterior of the treatment effect $\delta = \theta_1 - \theta_0$ is  $\Pi_\delta(\cdot \mid  D)$.  Investigators report evidence of a positive effect if $\Pi_\delta \left((0, \infty]\mid D\right) \geq 0.9$. 

\textbf{Simulation model and scenarios:} To discuss  asymptotic features of the Q-approximation procedure, we consider a sequence of scenarios $\omega_{n}$ in which the treatment effect is a function of the sample size. For a sample size $n$, we set $\omega_{n} = \left(\omega_{0,n}, \phantom, \omega_{1,n} \right)$, where $\omega_{0,n} = 0.4$  and $\omega_{1,n} = \omega_0 + \frac{a}{\sqrt{n}}$. In our sequence of scenarios $n_1 = n_2$ and $n = n_1 + n_2 \rightarrow \infty$. The simulation model for a specific sample size $n$ is
\bne \label{model:ex2_simulation_outcome}
Y_i \mid A_i = k \overset{iid}{\sim} \mbox{Ber}\left(\omega_{k,n}\right),\; k = 0,1.
\ene 
We set $a = 2.1$, which ensures that the power converges close to 80\% as $n\to\infty$. 

\textbf{Operating characteristic:} The OC of interest is the probability of reporting evidence of positive effects. Therefore,
\bne \label{eq:ex2_h}
h(D) = \mathbbm{1} \left\{ \Pi_\delta((0, \infty]\mid D)\geq 0.9 \right\}.
\ene

\begin{algorithm}[t]

\caption{MC-approximation and Q-approximation, Example 2. OC: probability of reporting a positive result.}

\vspace{0.5em}

\begin{subalgorithm}[H]{\textwidth}

\caption{\textbf{MC-approximation}}

\label{alg:MC_ex2}

\textbf{Input:} $n$, prior $\pi_\theta(\cdot)$, $\omega_n$, M, R
\begin{algorithmic}[1]
\For{ replicate $r$ in $1$:$R$ } 
    \State Sample $D^{(r)} = (Y_{i}, A_i)_{i\leq n}$, where $Y_{i} \mid A_i = k \overset{\text{iid}}{\sim} \text{Ber}(\omega_{n,k})$, $k = 0,1$.
    
    \For{$m$ in  $1$:$M$}
        \State Sample $\theta^{(r,m)}$ from $\Pi_\theta\left(\cdot  \mid D=D^{(r)}\right)$.
    \EndFor
    \State Compute $h \left( D^{(r)} \right) = \mathbbm{1}\left\{\left(\frac{1}{M} \sum_{m = 1}^M \mathbbm{1}\left\{\theta_1^{(r,m)} - \theta_0^{(r,m)} \geq 0\right\}\right) \geq 0.9\right\}$.
\EndFor

\end{algorithmic}

\textbf{Output:} $\frac{1}{R} \sum_{r = 1}^R h\left(D^{(r)}\right)$
    
\end{subalgorithm}

\vspace{1em}

\begin{subalgorithm}[H]{\textwidth}

\caption{\textbf{Q-approximation}}
\label{alg:Q_ex2}

\textbf{Input:} $n$, prior $\pi_\theta(\cdot)$, $\omega_n$, R
\begin{algorithmic}[1]
\For {replicate $r$ in $1$:$R$}

\State \multiline{
    \raggedright
    Generate the Q-likelihood $\widetilde{L} \left( \cdot; C^{(r)}, V \right)$, where $C^{(r)} = \bmat{C_0^{(r)} \\ C_1^{(r)}} \sim N\left(\begin{bmatrix}
    \omega_{0,n} \\
    \omega_{1,n} \\
    \end{bmatrix},
    \begin{bmatrix}
    \frac{\omega_{0,n}(1 - \omega_{0,n})}{n_0} & 0 \\
    0 & \frac{\omega_{1,n}(1 - \omega_{1,n})}{n_1}
    \end{bmatrix}
    \right)$ and $V = \begin{bmatrix}
    \frac{n_0}{\omega_{0,n}(1 - \omega_{0,n})} & 0 \\
    0 & \frac{n_1}{\omega_{1,n}(1 - \omega_{1,n})}
    \end{bmatrix}$.
}

\State 
\multiline{
    \raggedright
    Compute $h'\left( \widetilde{L} \left( \cdot; C^{(r)}, V \right) \right) = \mathbbm{1}\left\{\widetilde{\Pi}_\delta \left( (0, \infty]; C_\delta^{(r,P)}, V_\delta^{(P)} \right) > 0.9 \right\} = \mathbbm{1}\left\{ \left [1 - \Phi\left(0; C_\delta^{(r,P)}, V_\delta^{(P)^{-1}} \right)\right ] > 0.9 \right\}$, where $C_\delta^{(r,P)} = C_1^{(r)} - C_0^{(r)}$ and $V_\delta^{(P)^{-1}} = trace \left( V^{-1} \right)$.
}
\EndFor
\end{algorithmic}

\textbf{Output:} $\frac{1}{R} \sum_{r = 1}^{R} h'\left( \widetilde{L} \left( \cdot; C^{(r)}_{1:S}, V^{(r)}_{1:S} \right) \right)$
    
\end{subalgorithm}
    
\end{algorithm}

\textbf{MC-Approximation}: 
The key steps for a standard MC-approximation of this OC are described in Supplementary Algorithm~\ref{alg:MC_ex2} and are very similar to Algorithm \ref{alg:MC_ex1} in \textbf{Example 1}. With the conjugate Beta-Bernoulli model the posterior of $\theta_k$, for $k=0, 1$, is $\text{Beta} \left(\alpha_k + Y_{.,k}, \beta_k + (n_k - Y_{.,k})  \right)$ where $Y_{\cdot,k} = \sum_{i : A_i = k} Y_i$. The posterior probability $\Pi_\delta \left((0, \infty]\mid D^{(r)} \right)$ is approximated by Monte Carlo integration, sampling repeatedly from the posterior of $\theta$.

\textbf{Q-Approximation}: Supplementary Algorithm~\ref{alg:Q_ex2} outlines the key steps of the Q-approximation for the same OC. It differs from the MC-approximation in two aspects. First, instead of sampling datasets $D^{(r)}$ (alg \ref{alg:MC_ex2}, line 2), the Q-approximation algorithm samples Q-likelihoods $\widetilde{L} \left(\cdot; C^{(r)}, V \right)$ with random centers $C^{(r)}$ and fixed curvature $V$ (alg \ref{alg:Q_ex2}, line 2). Second, instead of computing posterior probabilities by Monte Carlo integration (alg \ref{alg:MC_ex2}, lines 3-6), the Q-approximation simply evaluates a Gaussian CDF (alg \ref{alg:Q_ex2}, line 3). 

The interpretation of the Q-likelihoods (alg \ref{alg:Q_ex2}, line 2) is the same as in Section \ref{section:single_stage_Q}. In particular, based on Considerations A and B, we have a Gaussian approximation of the likelihood $L \left( \theta ; D \right)$:
\begin{equation} \label{eq:gaussian_lik_ex2}
     \phi_2 \left( \theta; \hat{\theta}, J^{-1} \right) = \phi_2 \left(
    \begin{bmatrix} 
        \theta_0\\
        \theta_1
    \end{bmatrix}; 
    \begin{bmatrix} 
        \overline{Y}_0\\
        \overline{Y}_1\\
    \end{bmatrix},
    \begin{bmatrix}
        \frac{\bar{Y}_0\left(1 - \overline{Y}_0\right)}{n_0} & 0 \\
        0 & \frac{\overline{Y}_1 \left(1 - \overline{Y}_1\right)}{n_1}
    \end{bmatrix}
    \right),
\end{equation}
where $\bar{Y}_1, \bar{Y}_0$ are the  mean outcomes in the treatment and control arms. Then, Consideration C suggests approximating the sampling distribution of $\widetilde{L} \left( \cdot; \hat{\theta}, J \right)=\phi_2 \left( \cdot; \hat{\theta}, J^{-1} \right)$ with the distribution of $\widetilde{L} \left( \cdot; C^{(r)}, V \right)$, where $C^{(r)} \overset{iid}{\sim} f_{\omega, \mathcal{T}} = N\left( \theta^*, n^{-1} \mathcal{V}^* \right)$ (see expression \eqref{eq:mle_distribution}) and the curvature $V = n \mathcal{J}^*$. See Algorithm \ref{alg:Q_ex2} for explicit  expressions for $\theta^*$, $\mathcal{V}^*$, and $\mathcal{J}^*$.

Computing $h' \left( \widetilde{L} \left(\cdot; C^{(r)}, V \right) \right)$ (alg \ref{alg:Q_ex2}, line 3) is simple. This step of the algorithm provides  the posterior inference we would make on $\delta = \theta_1 - \theta_0$ if the likelihood function were $\widetilde{L} \left(\cdot; C^{(r)}, V \right)$ with a flat prior. 
In Algorithm \ref{alg:Q_ex2} the Q-posterior density of $\theta$ --- obtained by combining the prior and the Q-likelihood --- is  $\widetilde{\pi}_\theta \left( \theta; C^{(r,P)}, V^{(P)} \right) \propto \widetilde{L} \left( \theta; C^{(r)}, V \right)$, a bivariate Gaussian density with mean   $C^{(r, P)} = C^{(r)} = \left( C_0^{(r)}, C_1^{(r)} \right)^\top$ and variance $V^{(P)^{-1}} = V^{-1}$. Here, $C^{(P)}$ and $V^{(P)}$ denotes the center and curvature of the Q-posterior. In this example, the prior is flat so they coincide with the center and curvature of the Q-likelihood; in more complex settings (e.g. Examples 3 and 4) they will differ. 

The Q-posterior of $\theta$ implies that the Q-posterior density of $\delta = \theta_1 - \theta_0$ is Gaussian with mean $C^{(r,P)}_\delta = C_1^{(r)} - C_0^{(r)}$ and variance $V^{(P)^{-1}}_\delta = trace\left( V^{-1} \right)$. 
When the prior on $\theta$ is not uniform (i.e., $\alpha_k\neq 1$ or $\beta_k \neq 1$ for $k = 0,1$ in model \eqref{model:ex2_analysis}), Algorithm \ref{alg:Q_ex2} remains nearly identical,  with slightly adapted  expressions  for $C_\delta^{(r,P)}$ and $V_\delta^{(P)}$ (see Supplementary Section S1.1).

\textbf{Consistency of the Q-approximation 
:} 
Recall that we specified a sequence of scenarios $\omega_n$ with decreasing treatment effects, and the power $ \psi_n=\mathbb{E}_{g_{\omega_n, \mathcal{T}}} \left( h(D) \right)$ converges  as $n \to \infty$. The following proposition states that the Q-approximation of the power (the output of Algorithm \ref{alg:Q_ex2}) converges to the same limit when the number of replicates $R \to \infty$ and the sample size $n \to \infty$. Here we use the notation $\hat{\psi}_{Q,n}$ for the Q-approximation of the power, or more generally the estimate of the operating characteristic of interest, 
to emphasize its dependence on $n$.

\begin{proposition} \label{prop_ex2}
    In Example 2, as $R \to \infty$, we have $\hat{\psi}_{Q,n} \overset{p}{\to} \psi_{Q,n} = E_{f_{\omega_n, \mathcal{T}}} \left[ h' \left( \widetilde{L} \left( \cdot; C^{(r)}, V \right) \right) \right]$. Also, as $n \to \infty$, we have $\psi_{Q,n} - \psi_n \to 0$.
\end{proposition}

\noindent Here $\psi_{Q,n}$ is simply the Q-approximation of $\psi_n$ when the number of replicates $R$ diverges.

Corollary \ref{cor_ex2} indicates that a similar result holds for other OCs. The corollary focuses on posterior expectations of bounded functions $u$ over $\Theta$. 

\begin{corollary} \label{cor_ex2}
    In Example 2, consider the sequence of OCs $\psi_n = \mathbb{E}_{g_{\omega_n, \mathcal{T}}} \left[ h'(L(\cdot ; D)) \right]$ where $h'(L(\cdot ; D)) = \mathbb{E}_{\Pi_\theta(\cdot \mid D)} \left[ u(\theta) \right]$ for some bounded $u: \Theta \to \mathbb{R}$. Let 
    \begin{equation*}
        \hat{\psi}_{Q,n} = \frac{1}{R} \sum_{r=1}^R h' \left( \widetilde{L} \left( \cdot; C^{(r)}, V \right) \right),
    \end{equation*}
    where $\widetilde{L} \left( \cdot; C^{(r)}, V \right)$ is generated as in Algorithm \ref{alg:Q_ex2}. Then as $R \to \infty$, we have $\hat{\psi}_{Q,n} \overset{p}{\to} {\psi}_{Q,n} = \mathbb{E}_{f_{\omega_n, \mathcal{T}}} \left[ h' \left( \widetilde{L} \left( \cdot; C^{(r)}, V \right) \right) \right]$. And as $n \to \infty$ we have $ {\psi}_{Q,n} - \psi_n \to 0$.
\end{corollary}
\noindent The proofs of these results are in Supplementary Section S1.2.

Figure \ref{fig:ex2} summarizes the main steps of the Q-approximation in Example 2  for the scenario $\omega_{100} = (0.4, 0.61)$, where $n = 100$, $n_1 = n_0 = 50$.
\textbf{Panel A} shows, for a single dataset, the similarity  between the likelihood $L(\theta; D)$ and its Gaussian approximation $\widetilde{L} \left( \theta; \hat{\theta}, J  \right)$ from equation \eqref{eq:gaussian_lik_ex2}. For the control group, the blue curve is  $L \left( (\cdot, \hat{\theta}_1); D \right)$,  a function of $\theta_0$, and the red curve is  the approximation $\widetilde{L} \left( (\cdot, \hat{\theta}_1); \hat{\theta}, J \right)$.
\textbf{Panel B} compares the sampling distribution of the MLE $\hat{\theta}^{(r)}$ across Monte Carlo replicates to the asymptotic approximation $f_{\omega_{100}, \mathcal{T}}$ used to sample $C^{(r)}$ in the Q-approximation (alg \ref{alg:Q_ex2}, line 2). 
\textbf{Panel C} illustrates that when $n$ increases, the ratio between the observed information $J_{0,0} = \frac{n_0}{\bar{Y}_0 ( 1 - \bar{Y}_0 )}$ (equation \ref{eq:gaussian_lik_ex2}) and the curvature $V_{0,0} = n_0 \mathcal{J}_{0,0} = \frac{n_0}{\omega_0 ( 1 - \omega_0 )}$ (the first entry of the $V$ matrix in Algorithm \ref{alg:Q_ex2}, line 2) has a sampling distribution that concentrates around 1.
\textbf{Panel D} displays four replicates of the function $h'\left( \widetilde{L} \left( \cdot; C^{(r)}, V \right) \right) = \mathbbm{1}\left\{\widetilde{\Pi}_\delta \left( (0, \infty]; C_\delta^{(r,P)}, V^{(P)}_\delta \right) > 0.9 \right\}$ (alg \ref{alg:Q_ex2}, line 3). In two of these four replicates $\widetilde{\Pi}_\delta \left( (0, \infty]; C_\delta^{(r,P)}, V_\delta^{(P)} \right)$ is above the 0.9 threshold.
\textbf{Panel E} illustrates,  as $n=n_0+n_1$ increases, the Q-approximation $\hat{\psi}_{Q}$ (red) and the MC-approximation $\hat{\psi}_{MC}$ of the power (blue).
\textbf{Panel F} compares the accuracy of the Q-approximation (red) and the MC-approximation (blue) when we  fix the computing time budget (x-axis).  Note that in this toy example it is simple to compute exactly the power, allowing us to provide  accuracy summaries of the Q- and MC-approximations. When the computing time budget is small  $\hat{\psi}_{Q}$ has lower root mean squared error (RMSE) than $\hat{\psi}_{MC}$. For large computing time budgets $\hat{\psi}_{MC}$ is more accurate, as expected. Indeed $\hat{\psi}_{MC}$ is an unbiased and consistent estimate of the OC of interest. Interestingly, the Q-approximation RMSE is  around ~3\% when the computing time is so limited that
the MC-approximation algorithm cannot simulate even a single trial (left part of the panel, before the blue curve starts).

\subsection{Evaluating the accuracy of Q-approximations}

We describe a procedure for evaluating the accuracy of Q-approximations. When we consider many scenarios $\omega$, a practical strategy is to compare scenario-specific (i) Q-approximations and (ii)   MC-approximations with low accuracy, based on a few trial replicates per scenario. We keep the computing time of the MC-approximations similar to or lower than the computing time of Q-approximations. Importantly, MC-approximations are unbiased, but with a limited number of replicates they have substantial variability. By modeling the distribution of the differences between the Q-approximations and MC-approximations, we  summarize the accuracy of our Q-approximations. This approach, which we describe  in the next paragraph, can also be used when evaluating OCs across many designs (e.g., nearly identical protocols with different sample sizes).

Consider a set of scenarios indexed by $b \in \left\{ 1, \ldots, B \right\}$, with parameters $\omega_b$ and OCs $\psi_b = \mathbb{E}_{g_{\omega_b, \mathcal{T}}}\left[ h(D) \right]$. The accuracy of the Q-approximations can be assessed as follows:
\begin{enumerate}
    \item \raggedright \textbf{Compute  Q-approximations}.
    For each scenario $b$, compute $\hat{\psi}_{Q, b} = R_Q^{-1} \sum_{r=1}^{R_Q} h'\left( \tilde{L}\left( \cdot\, ; \, C^{(r)}_b, V_b \right) \right)$ (as in equation \eqref{eq:q_approximation}) using  $R_Q$ replicates. We compute $\hat{\sigma}^2_{Q, b}$, an estimate of the variance of $\hat{\psi}_{Q, b}$ 
    based on standard central limit theorem arguments.

    \item \textbf{Compute small-scale MC-approximations}.
    For each scenario $b$, compute $\hat{\psi}_{MC, b} = R_{MC}^{-1} \sum_{r=1}^{R_{MC}} h \left( D^{(r)} \right)$ (as in equation \eqref{eq:mc_standard})
    with a small number $R_{MC}$ of replicates. Here, $R_{MC}$ is chosen to make
    the computing time of the MC-approximations lower than or close to the computing time of the Q-approximations. We compute $\hat{\sigma}^2_{MC, b}$, an estimate of the variance of $\hat{\psi}_{MC, b}$. 

    \item \textbf{Model the discrepancies}.  The discrepancies $\hat{\Delta}_b = \hat{\psi}_{Q, b} - \hat{\psi}_{MC, b}$ can be analyzed using a random effects model, with $\hat{\Delta}_b \mid \Delta_b \overset{ind.}{\sim} N \left( \Delta_b, \hat{\sigma}^2_{Q, b} + \hat{\sigma}^2_{MC, b} \right)$ 
    and $\Delta_b \overset{iid}{\sim} N \left( \Delta, \tau^2 \right)$. Extensions of this model could be used, for example to allow the expected discrepancies $\Delta_b$ to vary smoothly as we change the parameters of our scenario $\omega_b$. We compute Bayesian estimates $\hat{\Delta}$ and $\hat{\tau}^2$ 
    for this model using weakly informative priors \citep{rover_weakly_2021}.
\end{enumerate}
The estimates $\hat{\Delta}$ and $\hat{\tau}^2$ summarize the mean and variance of the expected discrepancies $\Delta_b = \psi_{Q, b} - \psi_{b}$ across scenarios. Small values of $\hat{\Delta}$ and $\hat{\tau}^2$ suggest that the Q-approximations are accurate.

We applied this procedure to Example 2, evaluating Q-approximation accuracy across $B = 96$ scenarios with treatment effects ($\omega_n = ( 0.4, \phantom{.} 0.4 + 2.1 \cdot n^{-1/2} )$) and sample sizes ($n \in \left\{ 50, 60, \dots, 990, 1000 \right\}$). The OC of interest was the power, the expectation of expression \eqref{eq:ex2_h} under model \eqref{model:ex2_simulation_outcome}. In this example $R_Q = 160,000$ and $R_{MC} = 100$, yielding nearly identical computing times for the MC- and Q-approximations. The average standard errors were $B^{-1} \sum_b \hat{\sigma}_{Q,b} = 0.1\%$ for the Q-approximations and $B^{-1} \sum_b \hat{\sigma}_{MC,b} = 4.0\%$ for the MC-approximations. We then estimated the mean  discrepancy,  $\hat{\Delta} = 0.27 \%$  and computed the estimate $\hat{\tau} = 0.94 \%$. Since in this toy example the  power $\psi_b$ can be computed exactly, we also evaluated the differences $\hat{\psi}_{Q, b} - \psi_b$, which had a mean across scenarios of $0.32\%$ and standard deviation of $0.84\%$. In this example, the variances of $\hat{\psi}_{Q,b}$, $b=1,\ldots, B$,  are close to zero, so — as expected — the mean and standard deviation of  the vector $(\hat{\psi}_{Q,b} - \psi_b;\; b=1,\ldots,B)$ are similar to the estimates $\hat{\Delta}$ and $\hat{\tau}$.

\subsection{Q-approximations with  multi-stage designs} \label{section:multi_stage_Q}

The Q-approximation introduced in Section \ref{section:single_stage_Q}  can be extended to multi-stage adaptive designs. We  consider a multi-stage design with $S$ interim analyses (IAs).  Each IA may trigger adaptations such as early stopping for futility \citep{ventz_use_2022} or updated randomization probabilities \citep{lee_bayesian_2010}.

Each stage  of the trial $s = 1, \dots, S,$ generates  data $D_s$ for $n_{\cdot,s}$ patients. The decision made at IA $s$ is based on the cumulative likelihood $L(\cdot; D_{1:s})$, where $D_{1:s} = (D_1, \dots, D_s)$. We assume that patient outcomes $Y_i$ are independent given treatment assignment and covariates, which implies that the cumulative likelihood can be factorized
\begin{equation} \label{eq:multi_stage_decomposition}
L(\theta;D_{1:s}) = \prod_{t = 1}^s L(\theta; D_t) = L(\theta; D_{1:(s-1)}) \cdot L(\theta; D_s),
\end{equation}
where $L(\cdot; D_s)$ is the stage-specific likelihood based on  the data collected during stage $s$. 

We can generate cumulative Q-likelihoods $\widetilde{L} \left(\cdot; C_{1:s}, V_{1:s}  \right)$ with  distribution similar to $L(\cdot; D_{1:s})$ by multiplying stage-specific Q-likelihoods.  In particular, as in Section \ref{section:single_stage_Q}, for each stage $t$ we  generate a stage-specific Q-likelihood $\widetilde{L} \left( \cdot; C_t, V_t  \right)$ with  distribution similar to $L(\cdot; D_t)$. These stage-specific Q-likelihoods  are multiplied, mimicking the factorization in \eqref{eq:multi_stage_decomposition}, to obtain cumulative Q-likelihoods:
\begin{equation}  \label{eq:lik_cum_approx}
    \widetilde{L} \left( \theta; C_{1:s}, V_{1:s} \right) = \prod_{t = 1}^s \widetilde{L} \left( \theta; C_t, V_t \right) = \prod_{t = 1}^s \phi_d \left( \theta; C_t, V^{{-1}}_t \right) = \phi_d \left( \theta; C_{1:s}, V^{{-1}}_{1:s} \right),
\end{equation}
where $C_{1:s} = V_{1:s}^{-1} \left( \sum_{t=1}^s V_t C_t \right)$ and $V_{1:s} = \sum_{t = 1}^s V_t$. Note that $\widetilde{L}(\cdot; C_{1:s}, V_{1:s})$ remains Gaussian.

The stage-specific Q-likelihoods $\widetilde{L}\left ( \cdot; C_t, V_t \right)$ in \eqref{eq:lik_cum_approx} are sampled in the same way as the Q-likelihood in Section \ref{section:single_stage_Q}, with $C_t \sim f_{\omega, \mathcal{T}_t}$. Here the distribution $f_{\omega, \mathcal{T}_t}$ and the curvature $V_t$ are sequentially computed,  based on the study protocol and $\widetilde{L} \left( \cdot; C_{1:(s-1)}, V_{1:(s-1)} \right)$, so that the stage-specific Q-likelihoods $(\widetilde{L} \left( \cdot; C_t, V_t \right);\; t=1,\ldots,S)$ have a joint distribution similar to $\left(L \left(\cdot; D_t \right);\; t=1,\ldots,S\right)$. These sequential computations mimic the dependence between the $s$-th stage-specific $L(\cdot;D_s)$ and $L(\cdot;D_{1:(s-1)})$, which triggers  the IA decisions preceding the $s$-th stage of the trial.
In some cases, including the extension of Example 2 in the next paragraph,  $f_{\omega, \mathcal{T}_t}$ will remain the same across stages and replicates $r=1,\ldots,R$. In other cases, including designs with adaptive randomization such as Example 4 in Section 4, they will differ across stages and replicates $r=1,\ldots,R$.

We describe a trial  similar to our Example 2 (Section 2.3) but with a protocol that has $S$ IAs. For simplicity outcomes are assumed to be observed immediately after treatment. At each IA, the trial is stopped for futility if $\Pi_\delta \left( (0,1] \mid D_{1:s} \right) \leq \lambda_s$, where 
$0 <\lambda_1 < ... < \lambda_{S - 1} < \lambda_S = 0.9$. Therefore the final stage $S_{stop}\in \{1, \dots, S\} $ is  a random variable. The OC of interest is the early stopping probability, defined as $\mathbb{E}_{g_{\omega, \mathcal{T}}} \left( h(D) \right)$ with $ h(D) = \mathbbm{1} \left\{ S_{stop} < S \right\}$. The Q-approximation proceeds as follows. For each replicate $r = 1, \dots, R$, we sequentially generate stage-specific Q-likelihoods and cumulative Q-likelihoods to conduct IAs through the following steps.
    \begin{enumerate}[topsep=0pt]
        \item For each stage $s = 1, \dots, S,$ sample the stage-specific center $C_s^{(r)} \sim N \left( \theta^*, n^{-1}_{\cdot, \hspace{0.1em} s} \mathcal{V}^* \right)$ and compute the stage-specific curvature $V_s = n_s \mathcal{J}^*$. This step is identical to Algorithm \ref{alg:Q_ex2}, line 2, with the only exception that the stage-specific sample size $n_{\cdot, \hspace{0.1em} s}$
        replaces the sample size $n$.
        
        \item Combine the stage-specific Q-likelihoods to obtain the cumulative Q-likelihood
        $ \widetilde{L} \left( \cdot; C_{1:s}^{(r)}, V_{1:s} \right) = \phi_d \left( \cdot; C_{1:s}^{(r)}, V_{1:s}^{-1} \right), $ where $ C_{1:s}^{(r)} = V_{1:s}^{-1} \left( \sum_{t=1}^{s} V_{t} C_{t}^{(r)} \right)$ and $V_{1:s} = \sum_{t=1}^{s} V_{t}$.
        
        \item Mimic the IA using $\widetilde{L} \left( \cdot; C_{1:s}^{(r)}, V_{1:s} \right)$. 
        If $s = S$ or if $\widetilde{\Pi}_\delta \left( (0, 1] ; C^{(r,P)}_{1:s}, V^{(P)}_{1:s}\right) < \lambda_s$ (c.f. alg \ref{alg:Q_ex2}, line 3) we set $S^{(r)}_{stop} = s$,  otherwise we continue to stage $s+1$.
    \end{enumerate}
Finally, the Q-approximation of the early stopping probability is$\frac{1}{R} \sum_{r = 1}^R \mathbbm{1} \left\{S_{stop}^{(r)} < S\right\}.$

In this example, we consider a trial with fixed features --- in particular, the sample size and the randomization probabilities --- across stages. In Section \ref{section:BAR}, we consider a more general case and apply the Q-approximation to a design with Bayesian adaptive randomization (BAR) \citep{lee_bayesian_2010}.

\section{An RCT design that leverages external data} \label{section:Ventz}

We discuss Q-approximations for the multi-stage trial design of \citet{ventz_use_2022}, which uses external data for futility stopping, our \textbf{Example 3}. 

\subsection{Trial protocol and simulation scenarios}

\textbf{Trial protocol:}
The data collected during the trial are $D^{(t)} = \left(Y_i^{(t)}, X_{i1}^{(t)}, X_{i2}^{(t)}, A_i^{(t)} \right)_{i \leq n^{(t)}}$. Here $X_{i1}^{(t)}$ and $X_{i2}^{(t)}$ are binary pre-treatment covariates, $Y_i^{(t)}$ is a binary outcome,  and $n^{(t)}$ is the sample size.
The trial leverages an external dataset $D^{(e)} = \left( Y_i^{(e)}, X_{i1}^{(e)}, X_{i2}^{(e)}, A_i^{(e)} \right)_{i\leq n^{(e)}}$, with $A_i^{(e)} = 0 \; \text{ for } i=1,\ldots, n^{(e)}$ to improve interim decisions. In this section the superscripts $(t)$ and $(e)$ distinguish the randomized trial data from the external data.
During each stage $s= 1, \dots, S,$ the data $D_s^{(t)} =\left( Y_i^{(t)}, X_{i1}^{(t)}, X_{i2}^{(t)}, A_i^{(t)}\right)_{n^{(t)}_{\cdot,1:(s-1)} < i \leq n^{(t)}_{\cdot,1:s}}$ are collected from $n^{(t)}_{\cdot, \hspace{0.1em} s} = n^{(t)}_{\cdot,1:s} - n^{(t)}_{\cdot,1:(s-1)}$ patients (2:1 block-randomized,  treatment vs. control).  The stage-$s$ sample size is $n^{(t)}_{\cdot, \hspace{0.1em} s}=n^{(t)}_{0, \hspace{0.1em} s}+n^{(t)}_{1, \hspace{0.1em} s}$, including patients in  the treatment and control arms. For simplicity we assume that $n^{(t)}_{\cdot, \hspace{0.1em} s}$ is the same for all stages. 

At completion of each stage $s = 1, \dots, S-1$ we have an IA that could stop the trial early for futility. These interim decisions are based on a Bayesian logistic model,   
\begin{equation} \label{ex3_IA_model}
    \begin{aligned}
        & \text{Likelihood:} & Y_i^{(j)} \mid X_i^{(j)}, A_i^{(j)} \hspace{0.2em} &\overset{ind.}{\sim} \hspace{0.2em} \text{Ber} \left( F \left(\beta_0 + \beta_1 X_{i1}^{(j)} + \beta_2 X_{i2}^{(j)} +  \beta_3 A_i^{(j)} \right) \right), & \\
        & & & \hspace{-6em} \text{for all $i$ and $j = t, e$},\text{ and} \\
        & \text{Prior:} & \beta = [\beta_0, \beta_1, \beta_2, \beta_3]^\top &\sim N_4(0, \Sigma), &
    \end{aligned}
\end{equation}
where $F$ is the inverse logit function and $\Sigma = \text{\tt diag}_4(10)$. 
We call model \eqref{ex3_IA_model} the IA model. The trial is discontinued for futility at stage $s$ if $ \Pi_{Z} \left( \left( z_{1-\alpha}, \infty \right) \mid D^{(t)}_{1:s}, D^{(e)} \right) \leq \zeta$ for some fixed $\zeta \in (0, 1)$.  Here $\Pi_{Z} \left( \left( z_{1-\alpha}, \infty \right) \mid D^{(t)}_{1:s}, D^{(e)} \right)$ is the predictive probability, given the data at the end of stage $s$, that if the trial continues and enrolls $n^{(t)}$ patients (with no future IAs), the final analysis (FA) will report evidence of efficacy based on the $Z$-statistic \eqref{ex3_z_stat}. 
As discussed in \cite{ventz_use_2022}, $\Pi_{Z}$ marginalizes over the future patients' covariates $\left( X_{i1}^{(t)}, X_{i2}^{(t)} \right)_{i > n^{(t)}_{\cdot, 1:s}}$ using the empirical distribution of the previous patients' covariates $\left( X_{i1}^{(t)}, X_{i2}^{(t)} \right)_{i \leq n^{(t)}_{\cdot, 1:s}}$. See \cite{ventz_use_2022} for  details on the trial design.

If the study is not stopped early for futility, it  enrolls a total of $n^{(t)}$ patients, and the FA tests the null hypothesis of no treatment effect using the Z-statistic
\begin{equation} \label{ex3_z_stat}
    Z = \frac{ \hat{\gamma}_1 - \hat{\gamma}_0 }{ \sqrt{ \hat{\gamma}_{\cdot} \left(1 - \hat{\gamma}_{\cdot}\right) \left( \nicefrac{1}{n^{(t)}_{1, \hspace{0.1em} \cdot}} + \nicefrac{1}{ n^{(t)}_{0, \hspace{0.1em} \cdot}} \right) } },
\end{equation}
where $\hat{\gamma}_1 = \nicefrac{1}{n^{(t)}_{1, \cdot}} \sum_{i} Y_i^{(t)} A_i^{(t)}$, 
\hspace{0.1em} $\hat{\gamma}_0 = \nicefrac{1}{n^{(t)}_{0, \cdot}} \sum_{i} Y_i^{(t)} \left( 1 - A_i^{(t)} \right)$, \hspace{0.1em} $\hat{\gamma}_{\cdot} = \nicefrac{1}{n^{(t)}} \left[ n^{(t)}_{0, \hspace{0.1em} \cdot} \hat{\gamma}_0 + n^{(t)}_{1, \hspace{0.1em} \cdot} \hat{\gamma}_1 \right]$, and $n^{(t)}_{k, \hspace{0.1em} \cdot}$ is the final sample size of arm $k$. The FA does not involve external data and rejects the null hypothesis if $Z > z_{1-\alpha}$, the $(1 -\alpha)$ quantile of the standard normal distribution. 

The means $\hat{\gamma}_k$ in \eqref{ex3_z_stat} are  MLEs of the parameters $\gamma = \left( \gamma_0, \gamma_1 \right)$  with respect to the likelihood  model 
\begin{equation} \label{ex3_FA_model}
   Y_i^{(t)} \mid A_i^{(t)} = k \overset{ind.}{\sim} \text{Ber} \left( \gamma_k \right), \hspace{1em} \text{for $i = 1, \dots, n^{(t)}$ and $k = 0, 1$,}
\end{equation}
which is used for the FA.
We emphasize that in the previous sections we used $\theta$ to denote the parameters of a single likelihood model. In this example there are two likelihood models (the IA and FA models), with  parameters $\beta$ and $\gamma$.

\textbf{Simulation model and scenarios:} We simulate the data from a logistic model and
consider the same Scenarios 1-7 described in Table 1 of \cite{ventz_use_2022},
$$ Y_i^{(j)} \hspace{0.2em} \overset{ind.}{\sim} \hspace{0.2em} \mbox{Ber} \left( F \left(\omega_ 0 + \omega_1 X^{(j)}_{i1} + \omega_2 X^{(j)}_{i2} + \omega_3 X^{(j)}_{i3} + \omega_4 A_i^{(j)} \right) \right) \; \hspace{1em} \text{for all $i$ and $j = t, e$}. $$ 
The outcomes are observed immediately after randomization, and $X_i^{+^{(j)}} = \left( X_{i1}^{(j)}, X_{i2}^{(j)}, X_{i3}^{(j)} \right)$, for $j = t, e$, are discrete covariates with the same distributions as in \cite{ventz_use_2022}. Note that the covariate $X_{i3}$ is not available for data analyses. Therefore the IA model \eqref{ex3_IA_model} is misspecified when $\omega_3 \neq 0$ (Scenarios 5-7). In Scenarios 3, 4, 6, and 7 the distributions of the covariates in the trial and external populations are different. 

\textbf{Operating characteristics.}
We consider  three OCs: the type I error rate or power (depending on whether the treatment effect is zero or positive), the probability of an early stopping decision during the clinical trial, and the expected number of enrolled patients. 

\subsection{Algorithms to approximate the OCs}

\textbf{MC-approximation:} The Monte Carlo algorithm is described in \cite{ventz_use_2022}. It iteratively simulates the clinical trial, including all individual patient covariates and outcomes, as well as the MCMC computations to conduct each IA.

\begin{algorithm}[tb]

\setstretch{1}

\caption{Q-approximation, Example 3. OC: probability of reporting evidence of treatment effects.}
\label{alg:Q_ex3}

\textbf{Input:} 

protocol $\mathcal{T}$, scenario $\omega$,  and replicates $R$

\begin{algorithmic}[1]

\For { $r$ in $1$:$R$}

    \State \multiline{

    Generate the external Q-likelihood $\widetilde{L}^{IA} \left( \cdot; C_\beta^{(r, e)}, V_\beta^{(e)} \right)$.  
    {\it The distribution of $C_\beta^{(r, e)}$ and computation of $V_\beta^{(e)}$ are provided in Supplementary Section S2.3.2.}
    }

    \For {stage $s$ in $1$:$(S-1)$
    }

        \State \multiline{
        Jointly generate the 
        trial Q-likelihoods $\widetilde{L}^{IA} \left( \cdot; C_{\beta, \hspace{0.1em} s}^{(r, t)}, V_{\beta, \hspace{0.1em} s}^{(t)} \right)$
        and $\widetilde{L}^{FA} \left( \cdot; C_{\gamma, \hspace{0.1em} s}^{(r, t)}, V_{\gamma, \hspace{0.1em} s}^{(t)} \right)$. 
        {\it 
        See 
        Supplementary Section S2.3.3
        for the distribution of 
        $\left( C_{\beta, \hspace{0.1em} s}^{(r, t)}, C_{\gamma, \hspace{0.1em} s}^{(r, t)} \right)$ 
        and the value of 
        $(V_{\beta, \hspace{0.1em} s}^{(t)}, V_{\gamma, \hspace{0.1em} s}^{(t)})$.  
        }
        }

        \State \multiline{
        Compute the cumulative trial Q-likelihood $\widetilde{L}^{FA} \left(\cdot;  C_{\gamma, 1:s}^{(r,t)}, V_{\gamma,1:s}^{(t)} \right)$ and pooled Q-posterior
        $\widetilde{\pi}_\beta \left( \cdot; C_{\beta,1:s}^{(r,P)},
        V_{\beta,1:s}^{(P)} \right)$. 
        {\it See Eq. \eqref{eq:lik_cum_approx} and  Supplementary Section S2.3.4 for details.}
        }

        \SubStep{Conduct the IA: 
        }

        \State \multiline{
        Compute the Q-posterior predictive probability $\widetilde{\Pi}_Z^{(r,s)} \left( z_{1-\alpha} \right)$. {\it See  Eqs. \eqref{ex3_predictive_y_simple_binomial_sampled_beta} and \eqref{example_3_predictive_Z_eq}}.
        }

        \State \multiline{
        If $\widetilde{\Pi}_Z^{(r,s)} \left( z_{1-\alpha} \right) \leq \zeta$, stop the trial for futility and set $Z^{(r)}  = -\infty$.
        }
        
        \EndSubStep

    \EndFor

    \SubStep{ If the trial was not stopped for futility conduct the FA:}

    \State Compute the Q-approximated  Z-statistic $Z^{(r)}$. { \it See  Eq. \eqref{ex3_z_stat}.}


    \EndSubStep
      
\EndFor

\end{algorithmic}

\textbf{Output:} $\frac{1}{R} \sum_{r = 1}^{R} \mathds{1} \left\{ Z^{(r)} > z_{1-\alpha} \right\}.$

\end{algorithm}

\textbf{Q-approximation}: Algorithm \ref{alg:Q_ex3} summarizes our Q-approximations, which avoid sampling patient-level data and using MCMC. In this example, the trial design  involves two  models, the IA model \eqref{ex3_IA_model} and the FA model \eqref{ex3_FA_model}. 
Algorithm \ref{alg:Q_ex3} jointly generates the corresponding Q-likelihoods (alg \ref{alg:Q_ex3}, line 4).
The algorithm uses three key functions: 
\begin{enumerate}
    \item The first function (alg \ref{alg:Q_ex3}, line 2) generates  jointly 
    $\widetilde{L}^{IA} \left( \cdot ; C^{(r,e)}_\beta, V_\beta^{(e)} \right)$, an external IA Q-likelihood whose distribution mimics $L^{(IA)} \left( \cdot ; D^{(e)} \right)$, the IA-model likelihood based only on external data. 
    The Q-likelihood is Gaussian, with a normally distributed center $C^{(r,e)}_\beta$ and fixed curvature $V_\beta^{(e)}$ (see Supplementary Section S2.3.2 for formulae, based on standard asymptotic results). 

    \item The second function (alg \ref{alg:Q_ex3}, line 4) jointly generates $\widetilde{L}^{IA} \left( \cdot ; C^{(r,t)}_{\beta, \hspace{0.1em} s}, V^{(t)}_{\beta, \hspace{0.1em} s} \right)$ and $\widetilde{L}^{FA} \left( \cdot ; C^{(r,t)}_{\gamma, \hspace{0.1em} s}, V^{(t)}_{\gamma, \hspace{0.1em} s} \right)$, our IA-model and FA-model Q-likelihoods for stage $s$. These random functions jointly mimic the likelihood functions $L^{IA} \left( \cdot; D^{(t)}_s \right)$ and $L^{FA} \left( \cdot; D^{(t)}_s \right)$, which are based only on the stage-$s$ trial data $D^{(t)}_s$. The external data are not involved. The centers $C^{(r,t)}_{\beta, \hspace{0.1em} s}$ and $C^{(r,t)}_{\gamma, \hspace{0.1em} s}$ mimic the MLEs of $\beta$ and $\gamma$ in models \eqref{ex3_IA_model} and \eqref{ex3_FA_model} respectively, using only $D^{(t)}_s$. They are sampled jointly using standard asymptotic arguments (see Supplementary Section S2.3.3). 

    \item The third function (alg \ref{alg:Q_ex3}, line 7) computes $\widetilde{\Pi}_Z^{(r,s)} \left( z_{1-\alpha} \right)$, the Q-posterior predictive probability at the end of stage $s$ that the FA will report a significant treatment effect if the trial proceeds without further IAs. Since final testing is based on the FA model's sufficient statistics, with contributions from completed and future trial stages, this function has two inputs. The first one is the stage-specific Q-likelihoods $\widetilde{L}^{FA} \left( \cdot; C_{\gamma,s'}^{(r,t)}, V_{\gamma,s'}^{(t)} \right),\;\;s'=1,\ldots,s 
    $, which provides the 
      stage-specific 
    sufficient statistics 
    for the FA model
     (see Supplementary Section S2.3.4). The second one is the Q-posterior $\widetilde{\pi}_\beta \left( \cdot; C_{\beta,1:s}^{(r,P)}, V_{\beta,1:s}^{(P)} \right) \propto \widetilde{L}^{IA} \left( \cdot ; C^{(r,e)}_\beta, V_\beta^{(e)} \right) \times \prod_{s' = 1}^s \widetilde{L}^{IA} \left( \cdot; C_{\beta,s'}^{(r,t)}, V_{\beta,s'}^{(t)} \right) \times \pi_\beta(\cdot)$, which is used to predict the sufficient statistics during future stages $s+1, \dots, S$ (equation \eqref{ex3_predictive_y_simple_binomial_sampled_beta}).

    To compute $\widetilde{\Pi}_Z^{(r,s)} \left(\cdot \right)$, we repeat the following steps for $m = 1, \dots, M$:
    \begin{enumerate}
        \item \textit{Sample future arm-specific  responses.}
        Generate $\beta^{(r,m)} \sim \widetilde{\pi}_\beta \left( \cdot; C_{\beta, 1:s}^{(r,P)}, V_{\beta, 1:s}^{(P)} \right)$ and for $k = 0,1$ compute $$\gamma^{(r,m)}_k = \sum_{x \in \mathcal{X}} \hspace{0.2em} \tilde{p}_{x,1:s}^{(r,t)} \cdot F \left( \beta^{(r,m)}_0 + \beta^{(r,m)}_1 \mathds{1} \{ x_1 = 1 \} +  \beta^{(r,m)}_2 \mathds{1} \{ x_2 = 1 \} +  \beta^{(r,m)}_3 \mathds{1} \{ k = 1 \} \right),$$ where $\tilde{p}_{x,1:s}^{(r,t)}$ are weights identical to the empirical distribution of the individual profiles during stages $1:s$ (as per \cite{ventz_use_2022} protocol; see Supplementary Section S2.3.5). Then, for $k = 0, 1$, we sample
        \begin{equation} \label{ex3_predictive_y_simple_binomial_sampled_beta}
            \bar{y}^{(r,m)}_{k,(s+1):S} \sim \frac{1}{n^{(t)}_{k, (s+1):S}}  Bin\left( n^{(t)}_{k, (s+1):S}, \hspace{0.2em} \gamma^{(r,m)}_k \right).
        \end{equation}

        \item \textit{Compute the future Z-statistic}. Compute
        \begin{equation} \label{example_3_predictive_Z_eq}
            Z^{(r,m)} = \frac{ \tilde{\gamma}_1^{(r,m)}  - \tilde{\gamma}_0^{(r,m)} }{\sqrt{ \tilde{\gamma}_\cdot^{(r,m)} \left( 1 - \tilde{\gamma}_\cdot^{(r,m)} \right) \left( \left( n^{(t)}_{1,\cdot} \right)^{-1} + \left( n^{(t)}_{0,\cdot} \right)^{-1} \right) }},
        \end{equation}
        where $\tilde{\gamma}^{(r,m)}_k$ is a weighted average of future outcomes $\bar{y}^{(r,m)}_{k, \hspace{0.1em} (s+1):S}$ 
        from step (a) and observed outcomes $ \bar{y}^{(r)}_{k, \hspace{0.1em} 1:s} = C_{\gamma_k, 1:s}^{(r,t)}$ from $\widetilde{L}^{FA} \left( \cdot; C_{\gamma,1:s}^{(r,t)}, V_{\gamma,1:s}^{(t)} \right)= \prod_{s' = 1}^s \widetilde{L}^{FA} \left( \cdot; C_{\gamma,s'}^{(r,t)}, V_{\gamma,s'}^{(t)} \right)$. See Supplementary equation S2.3.22.
    \end{enumerate}

    Finally, $\widetilde{\Pi}_Z^{(r,s)} \left(\cdot \right) = M^{-1} \sum_{m=1}^M \mathds{1} \left\{ Z^{(r,m)} \in \cdot \right\}$. 
\end{enumerate}
Computing $\widetilde{\Pi}_Z^{(r,s)} \left(\cdot \right)$ does not require simulations of individual patient covariates or outcomes.

\subsection{OC estimates}

As in \cite{ventz_use_2022}, we set $n = 182$ and we have 4 IAs, after the enrollment of $36, \hspace{0.2em} 72, \hspace{0.2em} 108, \text{ and} \hspace{0.2em} 144$ patients respectively. Table \ref{table:ex_3} shows that the Q-approximations in this example closely agree with the MC-approximations of the OCs. For example, for the Type I error rate and power the difference between estimated OCs $|\hat{\psi}_{MC} - \hat{\psi}_Q|$ range from 0.03\% to 1.32\% across scenarios (Supplementary Table S2 includes the  confidence intervals for ${\psi}_{MC} - {\psi}_Q$).  
Across scenarios, the Q-approximation is approximately 1,700 times faster than the MC-approximation. Both Q-approximations and MC-approximations  involved  10,000 replicates per scenario.

\begin{table}[!htbp]
\centering
\resizebox{\textwidth}{!}{
\begin{tabular}{l c c c c c p{3em} c c c c c}
\toprule
Scen. &
\multicolumn{5}{c}{No treatment effect} &
&
\multicolumn{5}{c}{Positive treatment effect} \\
\cmidrule(lr){3-6} \cmidrule(lr){9-12}
 & OC & $\hat{\psi}_{MC}$ & $\hat{\psi}_Q$
 & $\hat{\psi}_{MC}-\hat{\psi}_Q$ & $T_{MC}/T_Q$
 & 
 &
 OC & $\hat{\psi}_{MC}$ & $\hat{\psi}_Q$
 & $\hat{\psi}_{MC}-\hat{\psi}_Q$ & $T_{MC}/T_Q$ \\
\midrule

\multirow{3}{*}{1}
 & Type I ER \% & 2.93 & 2.22 & 0.71 & \multirow{3}{*}{1891.2}
 & & Power \% & 79.35 & 78.79 & 0.56 & \multirow{3}{*}{1761.8} \\
 & \% St. & 93.36 & 94.08 & -0.72 &
 & & \% St. & 11.37 & 12.34 & -0.97 & \\
 & ESS & 66.30 & 65.10 & 1.20 &
 & & ESS & 169.60 & 168.30 & 1.30 & \\

\midrule
\multirow{3}{*}{2}
 & Type I ER \% & 2.68 & 2.57 & 0.11 & \multirow{3}{*}{1907.2}
 & & Power \% & 76.05 & 75.34 & 0.71 & \multirow{3}{*}{1694.2} \\
 & \% St. & 93.77 & 93.64 & 0.13 &
 & & \% St. & 14.46 & 15.25 & -0.79 & \\
 & ESS & 65.30 & 65.40 & -0.10 &
 & & ESS & 166.20 & 165.30 & 0.90 & \\

\midrule
\multirow{3}{*}{3}
 & Type I ER \% & 2.97 & 2.83 & 0.14 & \multirow{3}{*}{1783.6}
 & & Power \% & 76.66 & 75.34 & 1.32 & \multirow{3}{*}{1709.5} \\
 & \% St. & 93.00 & 92.90 & 0.10 &
 & & \% St. & 13.92 & 15.26 & -1.34 & \\
 & ESS & 66.30 & 66.70 & -0.40 &
 & & ESS & 166.80 & 165.10 & 1.70 & \\

\midrule
\multirow{3}{*}{4}
 & Type I ER \% & 2.70 & 2.53 & 0.17 & \multirow{3}{*}{1757.9}
 & & Power \% & 76.40 & 75.38 & 1.02 & \multirow{3}{*}{1750.1} \\
 & \% St. & 93.24 & 93.76 & -0.52 &
 & & \% St. & 14.05 & 15.15 & -1.10 & \\
 & ESS & 66.40 & 65.50 & 0.90 &
 & & ESS & 166.80 & 165.30 & 1.50 & \\

\midrule
\multirow{3}{*}{5}
 & Type I ER \% & 2.58 & 2.64 & -0.06 & \multirow{3}{*}{1745.8}
 & & Power \% & 72.66 & 71.69 & 0.97 & \multirow{3}{*}{1681.5} \\
 & \% St. & 93.51 & 93.90 & -0.39 &
 & & \% St. & 16.87 & 17.88 & -1.01 & \\
 & ESS & 66.30 & 65.30 & 1.00 &
 & & ESS & 163.70 & 162.80 & 0.90 & \\

\midrule
\multirow{3}{*}{6}
 & Type I ER \% & 3.23 & 3.11 & 0.12 & \multirow{3}{*}{1749.1}
 & & Power \% & 73.91 & 73.39 & 0.52 & \multirow{3}{*}{1713.5} \\
 & \% St. & 90.71 & 90.88 & -0.17 &
 & & \% St. & 12.88 & 13.69 & -0.81 & \\
 & ESS & 76.70 & 77.00 & -0.30 &
 & & ESS & 169.20 & 168.20 & 1.00 & \\

\midrule
\multirow{3}{*}{7}
 & Type I ER \% & 2.20 & 2.17 & 0.03 & \multirow{3}{*}{1797.3}
 & & Power \% & 68.06 & 67.45 & 0.61 & \multirow{3}{*}{1704.4} \\
 & \% St. & 95.33 & 95.55 & -0.22 &
 & & \% St. & 23.53 & 23.76 & -0.23 & \\
 & ESS & 57.20 & 56.40 & 0.80 &
 & & ESS & 154.90 & 154.50 & 0.40 & \\

\bottomrule
\end{tabular}
}
\caption{MC-approximation ($\hat{\psi}_{MC}$) and Q-approximation ($\hat{\psi}_Q$) of  three OCs:  Type I error rate (ER)/power (in \%),  scenario-specific likelihood  of  stopping the trial for fultility (\% St.), and expected sample size (ESS). The table  reports the difference $\hat{\psi}_{MC}-\hat{\psi}_Q$ and the ratio $T_{MC}/T_{Q}$ between the  computing time (in hours) required to perform 10000 MC-approximation replicates and 10000 Q-approximation replicates. MC computations and scenarios are the same as in \cite{ventz_use_2022}.}
\label{table:ex_3}
\end{table}

\section{A multi-arm RCT with Bayesian adaptive randomization} \label{section:BAR}

We discuss Q-approximations for a Bayesian adaptive randomization (BAR) design proposed by  \cite{lee_bayesian_2010}, our \textbf{Example 4}. 
 
\subsection{Trial protocol and simulation scenarios}

\textbf{Trial protocol:} The data collected during the study are
$
D = \{(Y_i, X_{i1}, X_{i2}, A_i)\}_{i \le n},
$
where $X_{i1}$ and $X_{i2}$ denote two binary biomarkers and
$A_i \in \{0,\dots, K-1 \}$ is the treatment assignment. In our comparisons $K = 4$. Each stage of the trial enrolls $n_s$ patients, and at the end of each stage $s$ the randomization probabilities $\rho_{k,s}^{(x)}$ that a patient with  biomarker profile $x = (x_1, x_2)$ receives treatment $k$ are updated. The probability  $\rho_{k,s}^{(x)}$ is proportional to the  posterior probability that treatment $k$ is optimal for patients with profile $x$. See \cite{thall_practical_2007} for details and a discussion. The randomization probabilities are computed using a Bayesian logistic regression model with interaction terms between treatments and biomarkers:
\begin{equation} \label{eq:protocol_model_ex4}
\begin{aligned}
    & \text{Likelihood:} & 
    Y_i \mid X_i &= x, A_i = k, \theta 
    \overset{ind.}{\sim} \mathrm{Ber}\!\left(F\!\left(\eta_k^{(x)}\right)\right), \\
&  &  \small
\eta^{(x)}_k
& = \theta_0
+ \theta_1\,\mathds{1}\{x_1 = 1\}
+ \theta_2\,\mathds{1}\{x_2 = 1\}
+ \sum_{j=1}^3 \theta_{2+j}\,\mathds{1}\{A_i = j\} +\;\;\;\;\;\;\;\;\;\;\;\;\;\;\;\;\;\;\;\;
\\[-1em]
& & & \hspace{2em} \;\;\;\;\;\;\;\; \;\;\;\;\sum_{j=1}^3 \Bigl(
    \theta_{5+j}\,\mathds{1}\{x_1 = 1\}
    + \theta_{8+j}\,\mathds{1}\{x_2 = 1\}
  \Bigr)\mathds{1}\{A_i = j\} 
 \\[0.3em]
    & \text{Prior:} & 
    \theta &= \left[\theta_0, \theta_1,...,\theta_{11} \right]^\top \sim N_{12}\!\left(0,\Sigma\right), 
    \quad \Sigma = \mathrm{diag}_{12}(10).
\end{aligned}
\end{equation}
Here $\eta^{(x)}_k$ is a linear function of $\theta$ that varies across biomarker profiles  and treatments. Let $\Pi_{\eta^{(x)}}(\cdot \mid D)$ denote the posterior distribution of $\eta^{(x)} = \left( \eta^{(x)}_0, \dots, \eta^{(x)}_{K-1} \right)$. At the beginning of the $s$-th stage  the available data are $D_{1:(s-1)} = \{(Y_i, X_{i1}, X_{i2}, A_i)\}_{i \le n_{1:(s-1)}}$, and patients with profile $x$ will be randomized to arm $k$ with probability
\begin{equation}\label{eq:superiority_probs}
\rho_{k,s}^{(x)} 
\propto 
\Pi_{\eta^{(x)}} \!\left(
\eta^{(x)}_{k}\; \ge\; \max_{j \ne k}\; \eta^{(x)}_{j}
\,\middle|\, D_{1:(s-1)}
\right),
\qquad k = 0,\ldots,K-1.
\end{equation}
At the end of stage $s$ the randomization probabilities for stage $s+1$, $\rho_{k,s+1}^{(x)}$, are computed for each arm $k$ and profile $x$.

\textbf{Simulation model and scenarios:}
We simulate data from the logistic model \eqref{eq:protocol_model_ex4} and focus on scenarios in which $X^+ = X$; hence, the analysis model is correctly specified. Additional details on the simulation model are provided in Supplementary Section S3.1. The design is adaptive, with randomization  probabilities $\rho_{k,s+1}^{(x)}$
that vary across stages and across simulation replicates. 
The outcomes are observed immediately after randomization.
We consider two scenarios similar to those reported in Table~4 of
\cite{lee_bayesian_2010} (see Supplementary Tables S3 and S4). In Scenario~1 all treatment effects are 0. In Scenario~2 all treatment effects are positive, but with a different magnitude for each biomarker profile.

\textbf{Operating characteristics:} We focus on OCs that summarize the distribution, across replicates of the trial, of the number of patients with a specific biomarker profile $x$ that are assigned to a specific arm $k$. As discussed in \cite{wathen_simulation_2017} and \cite{thall_statistical_2015} this distribution is a crucial component in the assessment of designs with adaptive randomization. In particular, we consider (1) the  arm-$k$ and profile-$x$ specific expected sample size (ESS) and (2) the standard deviation across trial replicates of the arm-$k$ and profile-$x$ specific sample size (SDSS).

\subsection{Algorithms to approximate operating characteristics}

\textbf{MC-approximation:}  The Monte Carlo algorithm is described in Supplementary Algorithm S2. It requires MCMC to sample from the posterior distribution of $\theta$ for each stage $s$ of every trial replicate $r$   (alg  S2, line 5). 

\textbf{Q-approximation:}
Algorithm~\ref{alg:q_ex_4} outlines the Q-approximation of the ESS. The trial is adaptive, with variations of the randomization probabilities. As a consequence, across stages, the parameters $\theta^*_s$, $\mathcal{V}^*_s$, and $\mathcal{J}^*_s$ necessary to generate the stage-specific Q-likelihood $\widetilde{L}\left(\cdot;C_{s}^{(r)}, V^{(r)}_s\right)$ change adaptively. This occurs because the KL divergence, expected curvature, and the expected Fisher information that define these three parameters (c.f. \ref{section:single_stage_Q}) change adaptively with the randomization probabilities (see Supplementary Equations (S3.3.2-S3.3.3)).
The remaining steps of the algorithm involve only linear transformations of the centers $C^{(r)}_{s}$ and curvatures $V^{(r)}_s$ (alg \ref{alg:q_ex_4}, lines 6-9). The Q-approximation of the probability that treatment $k$
is superior for patients with biomarker profile $x$, 
\[
\widetilde{\Pi}_{\eta^{(x)}} \!\left(
\eta^{(x)}_{k} \ge \max_{j \ne k} \eta^{(x)}_{j}
\,\middle|\, D_{1:s}
\right)
=
\widetilde{\Pi}_{\eta^{(x)}}\!\left(
\eta^{(x)}_{k,s} - \eta^{(x)}_{j,s} \ge 0
\;\; \forall j \neq k
\,\middle|\,
D_{1:s}
\right),
\]
is computed using a standard function
for multivariate normal orthant probabilities,  the
\texttt{pmvnorm} function in the \texttt{mvtnorm} R package
\citep{Genz2020} (alg \ref{alg:q_ex_4}, line 9). 
The Algorithm for the SDSS has the same structure, with a small modification discussed in Supplementary Section S3.4.

\begin{algorithm}
\caption{Q-approximation, Example 4. OC: arm $k$ and profile $x$ specific expected sample size.
}\label{alg:q_ex_4}
\textbf{Input:}

protocol $\mathcal{T}$, scenario $\omega$, and replicates $R$

\begin{algorithmic}[1]
\setstretch{1}
\For{$r = 1,\dots,R$}

    \State \multiline{
    Set initial randomization probabilities $\rho_{k, 1}^{(r,x)} \propto 1$ for all arms $k$ and profiles $x$.
    }

    \For{$s = 1,\dots,S-1$}
        \State \multiline{Compute $\theta_s^{(*,r)},  \mathcal{J}_s^{(*,r)}$ and $\mathcal{V}_s^{(*,r)}$ as functions of $\omega$ and the randomization probabilities $\rho_{k, s}^{(r,x)}$. 
        \textit{See Supplementary Section S3.3 for details.}}
        
        \State \multiline{Generate the stage-specific Q-likelihood $\widetilde{L}\left(\cdot; C_{s}^{(r)}, V_{s}^{(r)}\right)$,
        where $C_{s}^{(r)} \sim N_{}\left(\theta_s^{(*,r)}, n_s^{-1}\mathcal{V}_s^{(*,r)}\right)$ and $V_s^{(r)} = n_s \mathcal{J}_s^{(*,r)}$.}

        \State \multiline{Compute the cumulative Q-posterior $\widetilde{\pi}_{\theta}\left(\cdot; C_{1:s}^{(r,P)},V_{1:s}^{(r,P)}\right)$.
        \textit{Using Eq. \eqref{eq:lik_cum_approx} and conjugacy.}}

        \State \multiline{For each $x \in \mathcal{X}$ and $k = 0, \dots, K-1$ compute $\rho_{k,s+1}^{(r,x)}$ proportional to the the Q-posterior probability that treatment $k$ is superior for the patient profile $x$.
        \textit{See Supplementary Section S3.3  for details.}}
    \EndFor

        
\EndFor
\end{algorithmic}

\textbf{Output:} $\frac{1}{R} \sum_{r = 1}^{R} \sum_{s = 1}^S n_{s} \rho^{(r,x)}_{s, k} p_x$ for each $k = 0, \dots, K-1$ and $x \in \mathcal{X}$, where $p_x$ is the simulation prevalence of the profile $x$.
\vspace{0.3em}

\end{algorithm}

\subsection{A comparison of MC-approximations and Q-approximations}\label{sec:Results BAR}

We have a total sample size of $n = 160$, with patients initially randomized equally to the $K$ arms. We considered six different stage-specific sample sizes $n_s$, varying from 2 to 40 patients. The simulation scenarios are described in Supplementary Table~S3.

The results are summarized in Supplementary Figure~S1. In particular, \textbf{Panel A} shows that the Q-approximation and the MC-approximation produce very similar estimates of the OCs ESS and SDSS. The discrepancies between approximations remain nearly identical when we vary $n_s$. In particular, the Q-approximation provides results comparable to the MC-approximation even when $n_s = 2$, that is, when the stage-specific Q-likelihood refers to only two patients.

\textbf{Panel B} shows that the Q-approximation exhibits a small negative bias, slightly underestimating the arm-$k$ and profile-$x$ specific ESS and SDSS. The discrepancies between the MC-approximations and the Q-approximations in Scenario~1 are slightly more pronounced when $n_s$ is small and become less pronounced as $n_s$ increases. Overall, Q-approximations are between 267 and 706 times faster than MC-approximations across scenarios.

\section{Discussion}

We introduced \textit{Q-approximations}, a class of procedures for fast   approximation of the OCs of clinical trial designs. The proposed approach replaces Monte Carlo simulations of  clinical trials with simulations of quadratic approximations of the log-likelihood, called
Q-likelihoods. The centers and curvatures of the Q-likelihoods are simulated using standard asymptotic theory
and then transformed to construct all key trial quantities such as p-values and posterior probabilities.

This strategy makes the Q-approximation a general framework for conveniently evaluating the OCs of complex, likelihood-based trial designs, including multistage designs with early stopping (Section \ref{section:multi_stage_Q}), designs that leverage external data (Section \ref{section:Ventz}), and designs with adaptive randomization (Section \ref{section:BAR}). The framework is especially useful when broad design exploration is needed: Q-approximations can screen large design and scenario spaces, while standard Monte Carlo simulation can be reserved for final confirmation of the selected design.

Q-approximations can be orders of magnitude faster than Monte Carlo simulations. 
We reduce the computational budget  by sampling only the centers of the Q-likelihoods instead of repeatedly simulating patient profiles and outcomes. Furthermore, in many cases computations related to inference and prediction  are greatly simplified because the Q-likelihood is proportional to a Gaussian kernel, which allows closed-form evaluation of quantities that are otherwise intractable. This often eliminates the need for MCMC sampling, as in our Examples 3 and 4, leading to computation times hundreds to thousands of times shorter than Monte Carlo simulation.

A key feature of Q-approximations is that their accuracy can be assessed using Monte Carlo simulations. During the development of a clinical trial  protocol \citep[e.g.][]{kotecha_leveraging_2024}, investigators often need to explore hundreds of combinations of scenarios, candidate sample sizes, and other important parameters like the timing of IAs. In these cases, Q-approximations enable rapid exploration of a more comprehensive set of scenarios than would be practical to examine with standard Monte Carlo simulation. As discussed in Section 2.4, accuracy can be assessed with a modest computational budget using a small number of Monte Carlo replicates and a model to compare  MC- and Q-approximations. To summarize, key parameters of a trial design (e.g., sample size and critical thresholds for interim decisions) can be chosen based on comparisons informed by Q-approximations, and the accuracy of Q-approximations can quantified. Furthermore, the operating characteristics of the final design that will be implemented can be confirmed using standard Monte Carlo simulations.


We discussed asymptotic 
results that support 
the use of
Q-approximations. In particular
Proposition \ref{prop_ex2} and Corollary \ref{cor_ex2} show that, in the context of Example 2, the  accuracy of  Q-approximations is asymptotically guaranteed when we consider trials with large sample sizes. Our asymptotic results cover only simple trial designs in which most relevant OC can be easily computed without implementing approximation methods. The study of the asymptotic properties of Q-approximations could be extended to more complex  designs, such as our examples 3 and 4. Two considerations suggest that  convergence results similar to those derived for our example 2 might hold also in Examples 3 and 4. First, standard asymptotic results on logistic regression  \citep[Chapter 10.2]{van_der_vaart_asymptotic_2012} seem applicable to the study of stage-specific Q-likelihoods in Example 3 and Example 4. Second, we can hypothesize that if the  stage-specific approximation errors with Q-likelihood vanish, then also the    approximations of OCs combining multiple stage-specific ($s=1,\ldots,S$) Q-likelihoods present negligible errors. More broadly the characterization through asymptotic results of general sufficient  conditions for the accuracy of Q-approximations remains an open question.


\section*{Supplementary Material}
See the attached Supplementary Material file (PDF) for  notation tables, all derivations, and additional figures. See also the Supplementary Code Folder to reproduce the examples. 

\section*{Acknowledgments}
ChatGPT version 5.2 was used to revise and improve prose in some sections of the manuscript.

\section*{Funding}
We gratefully acknowledge support from the National Institutes of Health under grants R01LM013352 (LT and DS) and T32CA009337 (DS).

\section*{Conflict of Interest}
 We have no conflicts of interest to report.


\setstretch{1} 

\vspace{1em}

\bibliography{Draft/Approximation_paper.bib} 

\end{document}